\documentclass{aa}  

\usepackage{graphicx}

\usepackage{txfonts}
\usepackage{multirow}
\usepackage[version=4]{mhchem}
\usepackage{newtxtext,newtxmath}
\usepackage{tikz,xcolor, hyperref}

\definecolor{lime}{HTML}{A6CE39}
\DeclareRobustCommand{\orcidicon}{%
	\begin{tikzpicture}
	\draw[lime, fill=lime] (0,0) 
	circle [radius=0.16] 
	node[white] {{\fontfamily{qag}\selectfont \tiny ID}};
	\draw[white, fill=white] (-0.0625,0.095) 
	circle [radius=0.007];
	\end{tikzpicture}
	\hspace{-2mm}
}

\foreach \x in {A, ..., Z}{%
	\expandafter\xdef\csname orcid\x\endcsname{\noexpand\href{https://orcid.org/\csname orcidauthor\x\endcsname}{\noexpand\orcidicon}}
}

\begin{document}

   \title{Environmental history of filament galaxies}

   \subtitle{Stellar mass assembly and star-formation of filament galaxies}

   \authorrunning{D. Zakharova et al.}

      \author{D. Zakharova\inst{1}\orcidA{},
       G. De Lucia\inst{1,2}\orcidC{}
        B. Vulcani\inst{3}\orcidB{},
        F. Fontanot\inst{1,2}\orcidF{},
          L. Xie\inst{1,4}\orcidL{}
             }
    \institute{
     INAF – Osservatorio Astronomico di Trieste, Via Tiepolo 11, I-34131 Trieste, Italy
    \and
    IFPU - Institute for Fundamental Physics of the Universe, via Beirut 2, 34151, Trieste, Italy
    \and
INAF – Osservatorio astronomico di Padova, Vicolo dell’Osservatorio, 5, I-35122 Padova, Italy
\and 
Tianjin Normal University, Binshuixidao 393, 300387, Tianjin, China
    }

 
  \abstract
   { 
   Galaxy properties, such as stellar mass and star formation rate, correlate with their position within the cosmic web.  
   While galaxies are observed in a specific environment today, they may have experienced different environments in the past. This `environmental history', which is closely linked to pre-processing, is bound to leave an imprint on the observable and physical properties of galaxies.  
   In this work, we use the Galaxy Evolution and Assembly (GAEA) semi-analytic model and the magneto-hydrodynamic IllustrisTNG simulation to reconstruct the environmental histories of galaxies between $z=0$ and $z=4$ that today reside in filaments. Our goal is to understand how galaxy properties are related to their past environments, and to uncover the role of the cosmic web in shaping their present-day properties. This approach enables us to determine whether and when filamentary structures influence galaxy evolution. We find that filament galaxies at $z=0$ are a heterogeneous mix of populations with distinct environmental histories, with a clear dependency on the infall times into filaments. The vast majority of filament galaxies at $z=0$ have experienced group processing at some stage of their evolution, with only $\sim$20\% of galaxies remaining centrals throughout their life. For low-mass galaxies ($\rm 9 < \log_{10}(M_{star}/M_{sun}) < 10$), both GAEA and TNG100 confirm that the environmental effects on low-mass filament galaxies are primarily driven by group processing:  satellite populations have a stop growing stellar mass and exhibit elevated quenched fractions, whereas galaxies that remained centrals in filaments have properties that are similar to those of field galaxies. In contrast, massive galaxies ($\rm \log_{10}(M_{star}/M_{sun}) > 10$) do feel the effect of environment, regardless of being central or satellites. Massive galaxies that have never been satellites and that entered filaments more than 9 Gyr ago show accelerated stellar mass assembly and higher quenched fractions relative to the field, due to a higher frequency of merger events inside filaments, even at fixed stellar mass. Moreover, the most massive $\rm \log ((M_{star} / M_{sun}) > 11$) galaxies typically accreted onto filaments over 9 Gyr ago and have never become satellites within a larger halo, highlighting the role of filaments in building up the high-mass end of the galaxy population. These results demonstrate that filaments regulate galaxy evolution in a mass-dependent way: group environments regulate the evolution of low-mass galaxies, while filaments favour the growth of massive galaxies.
   
   }

   \keywords{Cosmology: large-scale structure of Universe, galaxies: evolution}

   \maketitle
%
\section{Introduction}
\label{sec:intro}

\indent The formation and evolution of galaxies are closely connected to the large-scale structure~(LSS) of the Universe. The latter can be expressed in terms of the spatial distribution of dark matter, gas, and galaxies, forming a web-like network known as the cosmic web~\citep{Bond+1996}. The presence of such large-scale structures is prominently revealed in wide-field surveys like 2dF~\citep{2df_survey} and SDSS~\citep{first_sdss_detection_of_lss}. 
Galaxies form and evolve within this cosmic web, experiencing different large-scale and local environments depending on their location within the LSS. They can reside in dense regions such as gravitationally bound clusters~(containing hundreds to thousands of galaxies), or in smaller groups, as well as in elongated structures that connect these systems, known as filaments or walls. In contrast, galaxies located in average or underdense regions are typically referred to as field galaxies. A growing number of studies, supported by observational evidence and theoretical predictions, indicate that galaxy properties exhibit systematic variations as a function of the environment, suggesting that the large-scale distribution plays a significant role in regulating galaxy evolution.  Galaxies residing in filaments are, on average, more massive and redder than field galaxies, and typically exhibit later morphological types, at least up to z<0.7~\citep{McNaught-Roberts+2014, Chen+2015, Guo+2015, Chen+2017, Malavasi+2017, Poudel+2017, Bonjean+2020, Laigle+2018, Kraljic+2018, Mahajan+2018, Luber+2019, Kotecha+2022, Hasan+2023, Zakharova+2023, OKane+2024}.
\par
Environmental effects on structure formation can be broadly divided into two categories: those that enhance halo growth and those that suppress it. On the one hand, filaments, as dense structures, are characterised by an enhanced reservoir of material for accretion compared to the field, leading to the formation of more massive haloes \citep{Lemson+1999, Mo+1996, Wang+2011}. On the other hand,  the cosmic web exerts a strong tidal field, which can suppress halo accretion and growth~\citep{Wang+2011, Musso+2018}. Several studies \citep{Hahn+2007, Lee+2017, Borzyszkowski+2017, Musso+2018} demonstrated that at the low-mass end, tidal fields prevent accretion of new material, while simultaneously allowing massive haloes in dense regions to continue accreting material. The properties of galaxies that reside within these haloes are affected by their environmental histories \citep{Croton+2007, De_Lucia+2012}, with massive galaxies gaining, on average, more stellar mass in dense environments \citep{Kraljic+2019}.
\par 
\par
Besides, galaxies in filaments may experience more frequent gravitational interactions than in the field~\citep{Kuutma+2017}, and increased major or minor mergers rates with gas-poor or gas-rich galaxies~\citep{mihos1996,Hopkins+2008}, harassment~\citep{Moore+1996}, or tidal interactions~\citep{Bekki+1998}. Finally, the filaments are full of groups or clusters which can be characterised by their specific environmental mechanisms. For instance, the interactions between the cold interstellar medium (ISM) of galaxies and the intergalactic medium (IGM) or intra-cluster medium (ICM). Depending on the strength or stage of interaction between ISM and ICM, this process is called ram-pressure stripping~\citep[RPS;][]{gunngott1972}, viscous stripping~\citep{Nulsen+1982} or thermal evaporation~\citep[]{Cowie+1977}. The influence of groups and clusters embedded in filaments can dominate the evolution of their member galaxies and leave a characteristic imprint on the overall galaxy population of filaments. This mixing of environmental effects may complicate the identification of the specific role of filaments in shaping galaxy properties, making the separation of group-driven processes a challenging task~\citep{Sarron+2019, Zakharova+2024}.
Besides, a few contradictory results on the role of filaments on galaxy properties have been reported. One is that environmental trends within filaments are negligible when controlling for stellar mass, as shown by several studies reporting little to no difference between filament and field galaxies~\citep{Alpaslan+2015, Eardley+2015, Perez+2024}. The key question is then to what extent the environment governs galaxy properties versus intrinsic factors such as stellar mass~\citep[see also][]{De_Lucia+2012, Kraljic+2018}. Finally, an additional complication is that galaxies do not remain in a single environment throughout their lifetime but can migrate between different environments, making it challenging to identify which stage played the dominant role in shaping their properties.
\par
In this paper, we aim to clarify the role of filaments in galaxy evolution while explicitly accounting for these two factors.  Our analysis is based on the IllustrisTNG simulation and the GAEA semi-analytic model, which together allow us to trace galaxy evolution across cosmic time within different environments. Specifically, we aim to reconstruct the environmental history of galaxies, i.e., the sequence of environments experienced by galaxies over cosmic time, and to distinguish between different evolutionary pathways of filament galaxies. In both models, we identify galaxies located in filaments that are also part of groups and those that have never resided in such environments. We focus on describing the full environmental history of galaxies with  $\log (\rm M_{star} / M_{sun}) > 9$  at z=0  that are located in filaments outside massive clusters $\log (\rm M_{halo} / M_{sun}) > 14$ . This approach enables us to investigate how key galaxy properties, such as stellar mass and star formation rate (SFR), are influenced by different environmental histories. Moreover, by controlling for stellar mass, we ensure that stellar mass-related processes do not drive the observed trends. Ultimately, this approach enables us to assess when and how galaxies in filaments begin to diverge from their field counterparts.  
\par
This paper is organised as follows. Section~\ref{sec:data} describes the details of the models considered: the hydrodynamical IllustrisTNG and the semi-analytical GAEA. Section~\ref{sec:method} explains the identification of galaxy environments at each snapshot and the method for reconstructing the galaxy environmental histories. Section~\ref{sec:res_env_hist_proc} presents the environmental histories of galaxies as predicted by both models and discusses the relevant timescales. Section~\ref{sec:res_props} explores how different environmental pathways affect galaxy properties, such as stellar mass and quenched fraction. In Section~\ref{sec:discussion}, we interpret the predicted trends, and finally, Section~\ref{sec:conclusion} summarises our findings.

\section{Theoretical models}
\label{sec:data}

We use two different models: the Galaxy Evolution and Assembly semi-analytic model~\citep[GAEA;][]{DeLucia+2014,Hirschman+2016,Xie+2020, Fontanot_etal_2020, De_Lucia+2024} and the magneto hydro-dynamical simulation IllustrisTNG-100~\citep[TNG100;][]{Illustris1, Illustris2, Illustris3, Illustris4, Illustris5}. Both models offer a robust framework for understanding galaxy formation and evolution, with comparable simulated volumes and resolutions, as described in detail below.

\subsection{The GAEA semi-analytic model}

We use predictions from the latest rendition of the GAlaxy Evolution and Assembly (GAEA) semi-analytic model~\citep{De_Lucia+2024} coupled with the Millennium II Simulation~\citep[MSII;][]{Boylan-Kolchin_etal}. GAEA\footnote{Information about the GAEA model and selected model predictions can be found at  https://sites.google.com/inaf.it/gaea} is based on the original model presented in \citet{DeLucia_and_Blaizot}, but includes a number of important updates that have been implemented over the years and that are summarised in~\citet[][hereafter, we refer to this model as GAEA2024]{De_Lucia+2024}. This latest rendition of the model provides an improved agreement with the observed distributions of specific star-fomation rates~(sSFR = SFR / $\rm M_{star}$) in the local Universe, as well as a quite good agreement with the observed passive fractions up to $z\sim 3$, making it an ideal tool to interpret the data considered in this work. 
    \par
    The MSII simulation, which represents the backbone of the model used in this study, follows 2,160$^3$ dark matter particles in a box of 100~Mpc$\,{h}^{-1}$ on a side, with cosmological parameters consistent with WMAP1 ($\Omega_\Lambda=0.75$,
    $\Omega_m=0.25$, $\Omega_b=0.045$, $n=1$, $\sigma_8=0.9$, and $H_0=73 \, {\rm km\,s^{-1}\,Mpc^{-1}}$). The resolution of the MSII simulation (the particle mass is $6.9\times10^6\,{\rm M_{sun}}\,{\rm h}^{-1}$) allows galaxies to be well resolved down to stellar masses of $\sim 10^8\,{\rm M}_{sun}$. Simulation outputs are stored at 68 snapshots, from z=127 to z=0. The cosmological parameters adopted for the MSII are not
    consistent with the latest parameters based on Planck data \cite{Planck_2016}, but the relatively small offsets are
    not expected to have a significant impact on model predictions~\citep[for a detailed comparison between MS-based and PMill-based predictions, see][]{Fontanot+2024}.

\subsection{The magnetohydrodynamical simulation IllustrisTNG-100}

    TNG100-1~\citep[hereafter TNG100;][]{Illustris1, Illustris2, Illustris3, Illustris4, Illustris5}  is a magneto hydro-dynamical~(MHD) simulation corresponding to a comoving volume of $75^{3}~h^{-3}$ Mpc$^{3}$ and a baryonic particle mass of $m_{b} \sim 1.4 \cdot 10^{6}~\rm{M}_{sun}$. TNG100 follows the formation and evolution of galaxies throughout cosmic time, including all relevant physical processes such us galactic
    winds from stellar feedback~\citep{Pillepich+2018},  and a treatment for black hole feedback~\citep{Weinberger+2017}. The 
    evolution of dark matter, cosmic gas, stars,
    and supermassive black holes is followed from redshift z=127 to z=0, and model results are stored at 100 snapshots. 
    \par
    The model adopts a $\Lambda$CDM cosmology with parameters consistent with  \cite{Planck_2016}: $\Omega_{\Lambda} = 0.6911$,  $\Omega_{m} = 0.3089$, $\Omega_{b} = 0.0486$, $H_{0} = 67.74$ km sec$^{-1}$ Mpc$^{-1}$, $\sigma_{8} = 0.8159$, $n_s$ = 0.9667.

\subsection{Models comparison}

The galaxy formation models used in this work, GAEA and IllustrisTNG, differ significantly in their treatment of galaxy evolution and environmental processes.  The GAEA2024 model combines dark matter only merger trees with analytic models for the description of baryonic processes, while TNG100 solves hydrodynamical equations for the co-evolution of dark and baryonic matter. Thus, TNG100, in principle, includes by construction a treatment of all types of interactions between galaxies and the IGM, including gas stripping (such as cosmic web stripping), tidal interactions, harassment, and the effects of large-scale cosmic web tidal fields. GAEA2024 includes an explicit treatment for partitioning the cold gas into its atomic and molecular components and for ram-pressure stripping of both the hot gas and cold gas reservoirs of satellite galaxies \citep{Xie+2017,Xie+2020}. It does not include any special treatment for galaxies in filaments~(in terms of, e.g.,  tidal fields or gas stripping). Both models account for assembly bias: the assembly of dark matter halos depends on properties other than halo mass, such as formation time. As demonstrated in earlier work, earlier-formed haloes are more clustered than later-formed haloes of similar mass~\citep{Gao+2005}, which is bound to have an impact on galaxy properties \citep{Croton+2007,Wang+2013}.

\section{Method}
\label{sec:method}
\subsection{Environment identification}
\label{sec:fils_idn}

To reconstruct the environmental histories of galaxies in the models in the redshift range 
0 < z < 4. We first identify the different environments within the simulated volume at each snapshot for all the most massive progenitors of  $\log (M_{\rm star} / M_\odot) > 9$ galaxies at z=0. In the following, a group is a system in which a central galaxy hosts at least one satellite at any snapshot, while a cluster corresponds to a group with a halo mass of $\log (\rm M_{halo} / M_{sun}) > 14$. Filamentary structures are identified separately in each snapshot. 
We consider snapshots 29–67 for GAEA2024 and 21–99 for TNG100,  corresponding to the redshift range 
0 < z < 4\footnote{The upper redshift limit is set to z=4 because the time intervals between snapshots in the Millennium Simulation II become too large at earlier times, limiting temporal resolution. The snapshot definition differs between the two models: GAEA (67 snapshots in total) has coarser time sampling, with redshift intervals of $\delta z \sim 0.3$ around z$\approx4$, while TNG100 (100 snapshots) provides finer sampling of $\delta z \sim 0.1-0.2$ around the same range.}. For each model and at each snapshot, filaments are identified employing the cosmic web finder DisPerSE \citep{Sousbie+2011, Sousbie_etal+2011} and adopting the procedure described below. 
\par

The galaxy filament system~(GFS) is built from the spatial distribution of galaxies. In order to determine the GFS with DisPerSE, we first consider the number of galaxies at each snapshot. Indeed, DisPerSE is sensitive to the number of galaxies used in filament identification: including more galaxies tends to reveal fainter filamentary structures. To eliminate the "bias" introduced by the increasing galaxy abundance over cosmic time, we adopt a fixed number of galaxies at each snapshot to construct the filaments: approximately 32,000 for GAEA and 19,000 for TNG100. These numbers correspond to the galaxy counts above a stellar mass threshold of $\log (M_{\rm star} / M_\odot) > 8$ at redshift $z = 4$. We adopt the lowest possible stellar mass threshold, set by the resolution limits of both models at $z = 4$, in order to maximise the number of galaxies available for identifying robust filaments at higher redshifts. As shown in \citet{Zakharova+2023}, selecting galaxies with $\rm \log (\rm M_{star} / M_{sun}) > 9$ is sufficient to robustly recover the cosmic web at $z\sim0$, ensuring that our mass cut at low redshifts does not lead to the loss of significant filamentary structures.

\begin{figure*}
    \centering
    \includegraphics[width=1\linewidth]{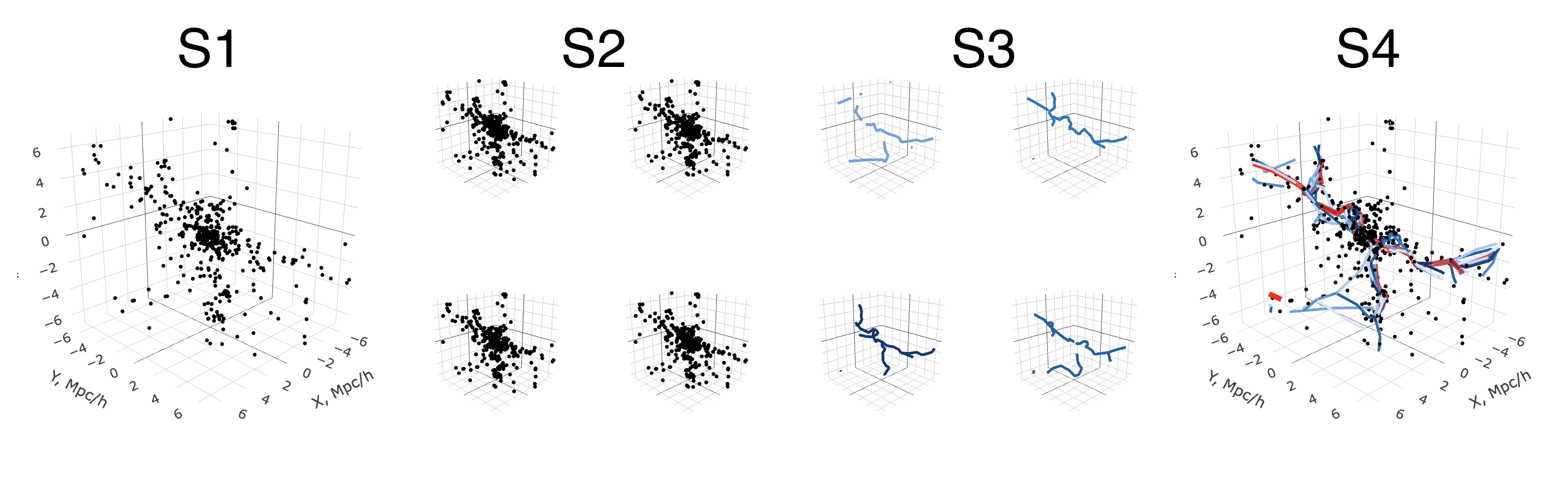}
    \caption{Four-step method for filament identification at each snapshot. The procedure is illustrated using a zoom-in region with side length 12 Mpc/$h$, centred on a massive halo with $\log_{10}(M_{\rm halo}/M_\odot)_{[z=0]} = 14.67$, at redshift $z=0$ in the GAEA2024 model. S1 shows the spatial distribution of all galaxies. S2 presents four random selections, each containing 85\% of the galaxy sample. S3 displays the filaments identified by DisPerSE for each corresponding random selection. The final panel, S4, shows all filaments across the four random selections separated by colour. The red lines represent the results of the commonly used approach of running DisPerSE once on all galaxies above the stellar mass threshold for the comparison.}
    \label{fig:method_desc}
\end{figure*}


Typically, filaments are identified by running DisPerSE once with carefully fine-tuned parameters \cite[e.g.,][]{Kraljic+2018, Sarron+2019, Singh+2020}. To enhance the robustness of the filament reconstruction, we instead identify the galaxy filamentary structure (GFS) at each snapshot through multiple realisations. Specifically, we perform 10 independent DisPerSE runs per snapshot, each based on a randomly selected subset of galaxies. Fig.~\ref{fig:method_desc} illustrates a typical region in the GAEA2024 simulated volume containing a massive halo. First, we extract the positions of all galaxies within the selected volume at a given snapshot (panel S1 in Fig.\ref{fig:method_desc}), using the stellar mass-selected samples mentioned above. Then, for each realisation, we randomly sample 85\% of the galaxies~(panel S2) and run DisPerSE with fixed parameters to extract the corresponding filament network~(panel S3). Filaments are identified using a persistence threshold of 5$\sigma$, selecting only the most prominent structures. We do not apply additional smoothing or rejection of short filaments. Instead, we rely on the repeated recovery of filamentary structures across realisations as an indicator of their robustness, avoiding further parameter tuning.
\par
Stacking the 10 sub-GFS realisations, we obtain robust filaments for each snapshot, as shown in Panel 4 of Fig.~\ref{fig:method_desc}. This process involves estimating the distance from each galaxy to each filament realisation. The median of these distances is used to obtain a robust measure of each galaxy’s proximity to the filament. This approach assumes that the 10 filament systems converge in the most densely populated regions. The main advantage of our method with respect to the standard approach lies in the mitigation of the effect of "flickering" filaments, e.g., the filament at the bottom of each box in the S3 panel that is not always identified. In a single DisPerSE run, flickering filaments disappear in adjacent snapshots. We further reduce the contribution of these features by running the process 10 times independently, thereby increasing the robustness of filament identification.
\par
We briefly discuss other possibilities for identifying the evolution of filaments through cosmic time. Rather than identifying independent GFS at each snapshot, we could (i) identify the filaments at each snapshot using the progenitors of z=0 filament members or (ii) use the descendants of z=4 filament members at subsequent snapshots. \\
The first approach would rely on the assumption that the progenitors of galaxies that today reside in filaments were always inside filaments, which may not be the case. \cite{Galarraga-Espinosa+2024} showed that the progenitors of present-day massive galaxies do not trace well the cosmic web at high redshifts. More importantly, such a method may allow us to characterise the evolution of filaments but not track the environmental history of galaxies.  
\par
The second method could provide a robust framework for our analysis. However, it assumes a rigid cosmic web, where new filaments do not form, collapse, or merge.  
Finally, both methods are limited in comparison to observations, where we lack access to either progenitors or descendants. In observational data, filaments are identified solely based on the spatial distribution of galaxies above a given stellar mass threshold.
\par
\begin{figure*}
    \centering
    \includegraphics[width=1\linewidth]{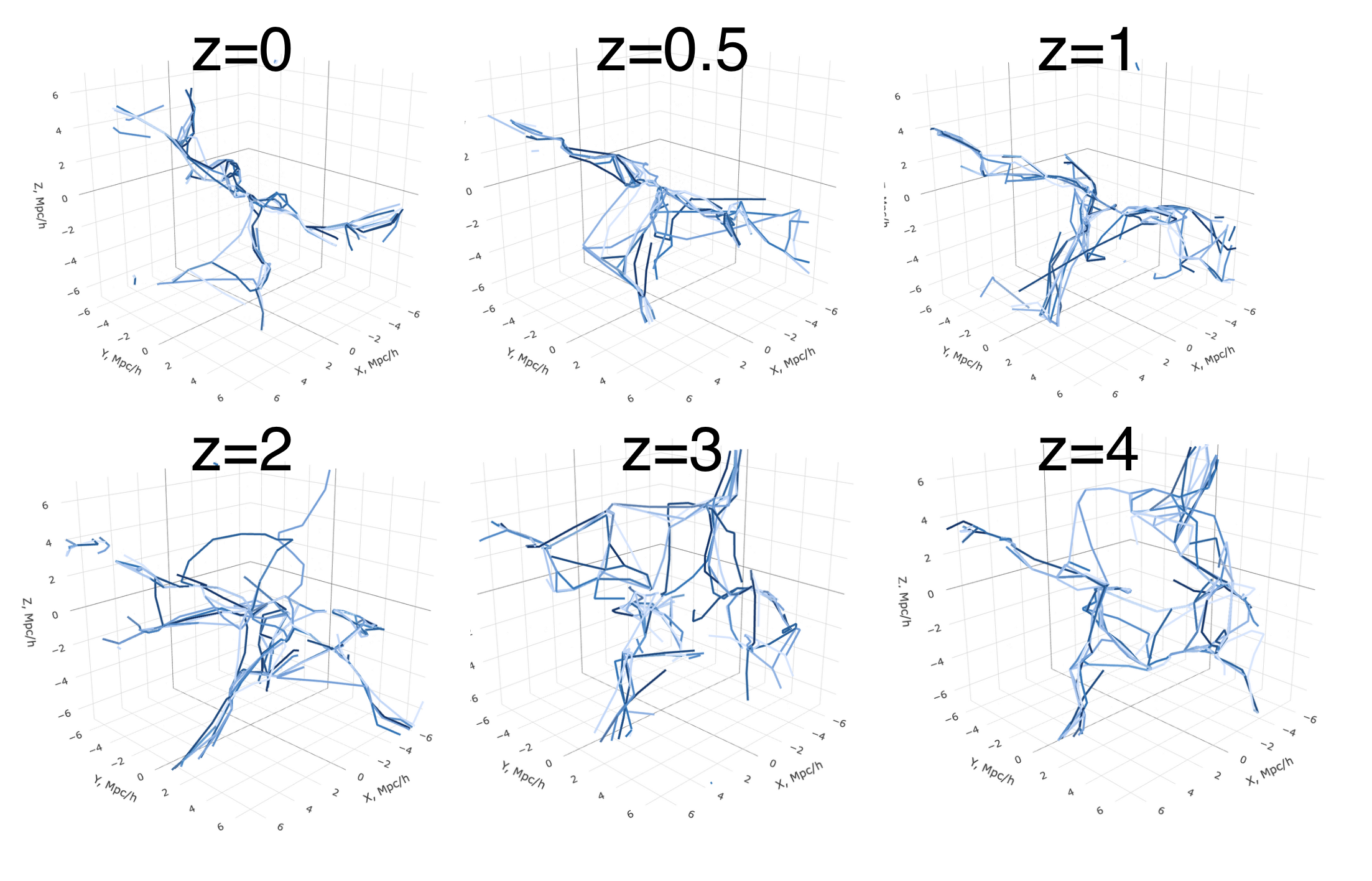}
    \caption{Evolution of the filamentary structure around the region that collapses into a $\rm \log (M_{halo} / M_{sun}) \approx 10^{14}$ cluster at $z=0$ in the GAEA2024. Different line colours indicate one of the 10 subGFS.}
    \label{fig:fils_evolution}
\end{figure*}

Finally, our method provides a consistent evolution of filaments within each simulation. An example is shown in Fig.~\ref{fig:fils_evolution}, illustrating the evolution of the filamentary network surrounding the region that forms a cluster $\rm \log (M_{halo} / M_{sun}) \approx 10^{14}$ between $z=0$ and $z=4$.

\subsection{Environmental history}
\label{sec:env_history_separation}

Having defined the environment at each snapshot, we now reconstruct the environmental histories of galaxies with $\rm \rm log (M_{\star}/M_{\odot}) > 9$ at $z=0$.  When tracing their progenitors back to $z=4$, we follow all galaxies with $\rm log (M_{\star}/M_{\odot}) > 7$, being aware that below this limit resolution effects become significant, motivating our choice of a higher mass cut for the main analysis. This ensures that only well-resolved galaxies are included, and that their progenitors can be reliably identified and tracked across cosmic time.
\par
We trace the main progenitors~(defined as the most massive stellar progenitor along the galaxy’s merger tree at each earlier snapshot) of each galaxy at z=0, and evaluate the following parameters at each snapshot for every progenitor:
\begin{itemize}
    \item Distance to the filament axis $\rm D_{\rm fil}$, defined as the median 3D distance to the nearest filament in each 10 subGFS estimated by bootstrapping galaxy positions;
    \item Type of the galaxy (central or satellite). GAEA2024 provides this parameter directly as an output. In TNG100, a galaxy is classified as central if its ID matches the halos \texttt{GroupFirstSub} (the first subhalo of the parent halo identified by \texttt{SubhaloGrNr}); otherwise, it is a satellite;
    \item Stellar mass. The TNG100 stellar mass measures all star particles gravitationally bound to the subhalo, while GAEA2024’s stellar mass is a model-based total stellar mass;
    \item Total host halo mass enclosed within $\rm R_{crit, 200}$, defined as the radius within which the average density is 200 times the $\rho_{crit}$. We refer to this halo mass as $\rm M_{\rm halo}$ hereafter;
    \item Star formation rate. GAEA2024 provides an averaged star-formation rate over $\sim 100$~Myr at high redshift and $\sim 350$~Myr at redshift below 0.8, while TNG100 provides only instantaneous star-formation rates, which correspond to the star-forming gas cells associated with the considered subhalo. The catalogues published in \cite{Donnari+2021} and \cite{Pillepich+2019_sfr} contain values averaged over specific time-scales, but only for galaxies with $>\sim$100 stellar particles. Since we include lower mass progenitors, not all of them have a defined average star formation rate at all snapshots.  Therefore, we use a different definition of the star-formation rates for models.
\end{itemize}
Additionally, for each galaxy at $z = 0$, we estimate the following parameters: (i) $\tau_{\rm fils}$ -- the lookback time when a galaxy first infalls onto a filament and is identified as a filament galaxy in at least the next two snapshots. This infall time is defined as the snapshot\footnote{We note that TNG100 have finer time steps than GAEA2024.}  where the galaxy approaches a filament at a comoving distance less than the filament radius, $\rm D_{\rm fil} < 1$ Mpc/h,\footnote{We neglect cases of filament backsplashes, even if they occur.}; (ii) $\tau_{\rm group~infall}$ -- the lookback time when a galaxy becomes for the first time a satellite of a group and remains a group member for the two subsequent snapshots. 
(iii) $\rm M_{70}$ -- the lookback time when a galaxy accumulated 70\% of its final stellar mass. Moreover, a galaxy is considered quenched if its specific star formation rate sSFR  = SFR / $\rm M_{star} < 10^{-11} yr ^{-1}$.
\par
\begin{figure}
    \centering
    \includegraphics[width=1\linewidth]{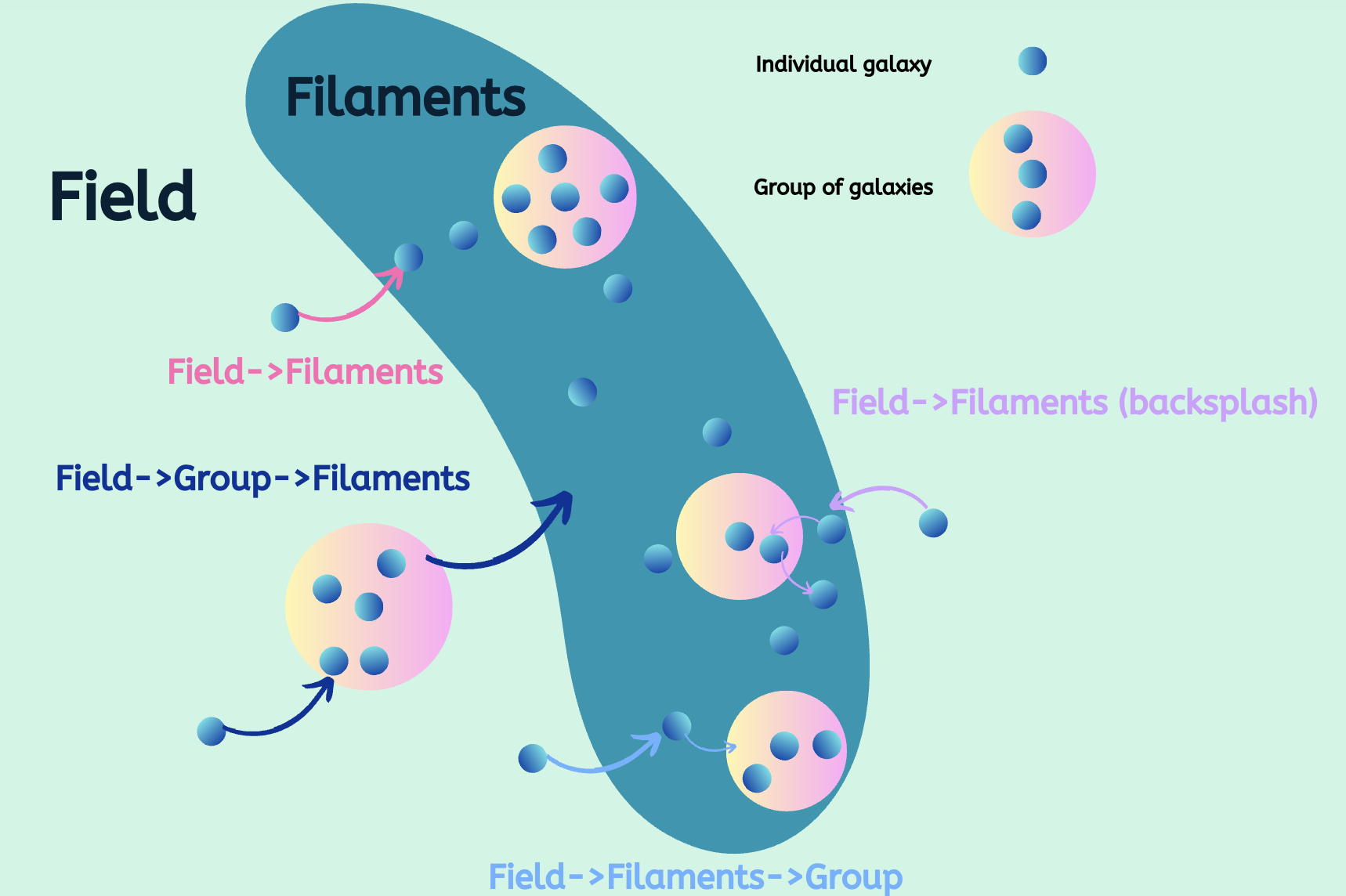}
    \caption{Cartoon representing filament galaxies' possible environmental history, as explained in Sec.~\ref{sec:env_history_separation}.  }
    \label{fig:env_branches}
\end{figure}


Using these quantities, we describe the environmental history of z=0 galaxies
that are located in filaments. From now on, we exclude filaments within massive clusters~($\log_{10}(\rm M_{\rm halo}/\rm M_{sun}) > 14$) because cluster galaxies experience a strong cluster-specific environment, which is not the focus of this study and would bias our results.
The GAEA2024 catalogue contains 5260  galaxies with $\log_{10}(\rm M_{\rm star}/\rm M_{sun}) > 10$ (from now on massive galaxies) and  10109 galaxies with $9 < \log_{10}(\rm M_{\rm star}/\rm M_{sun})< 10$ (from now on low-mass galaxies)  in filaments. The corresponding numbers for TNG100 are 2470 and 4610, respectively. Taking into account   $\tau_{\rm fil}$ and $\tau_{\rm group \, infall}$, we identify four populations of filament galaxies with different environmental histories as indicated in the cartoon shown in Fig.~\ref{fig:env_branches}:

\begin{itemize}
    \item \textbf{Field->Filaments~(FF)}: these are galaxies that fell from the field into the filaments as central galaxies and remain central down to z=0. This population of central galaxies can span the entire range of halo masses considered. We consider that this population has only experienced the filament environment. 
    \item \textbf{Field->Filaments [backsplash]~(FF-B)}: these are z=0 central galaxies that fell into filaments from the field, but that at some point in their history became satellites for at least two subsequent snapshots, i.e., backsplash population.\footnote{We do not distinguish cases where these galaxies interact with halos as satellites before or after their infall onto a filament infall.}
    \item \textbf{Field->Filaments->Group~(FFG)}: these are galaxies that fell from the field into filaments as central galaxies, but then become satellites of a group~($\gtrsim 10^{12},M_\odot$)  $\tau_{\rm group~infall}$ < $\tau_{\rm fils}$.
    \item \textbf{Field->Group->Filaments~(FGF)}: these are galaxies that fell into filaments as part of a group and stay in a group environment until z=0 ($\tau_{\rm group~infall}$ > $\tau_{\rm fils}$ ).
\end{itemize}
Finally,  we identify as field galaxies those central galaxies that, at any snapshot, are farther away than 1 Mpc/h  from any filament axis and that were never part of a group. This population constitutes our control sample and includes 5534 and 11860 massive and low-mass galaxies in GAEA2024, and 1377 and 4281 galaxies in TNG100.
\par
At this point, we can investigate the role of filaments by directly comparing field galaxies with the FF population, taking into account the time spent within the filaments. The FF-B category will highlight the role of backsplash galaxies, while the FFG or FGF pathways will emphasise the influence of the group environment on galaxy evolution.

\section{Environmental history of z=0 filament galaxies}
\label{sec:res_env_hist_proc}

\begin{figure}
    \centering
    \includegraphics[width=1\linewidth]{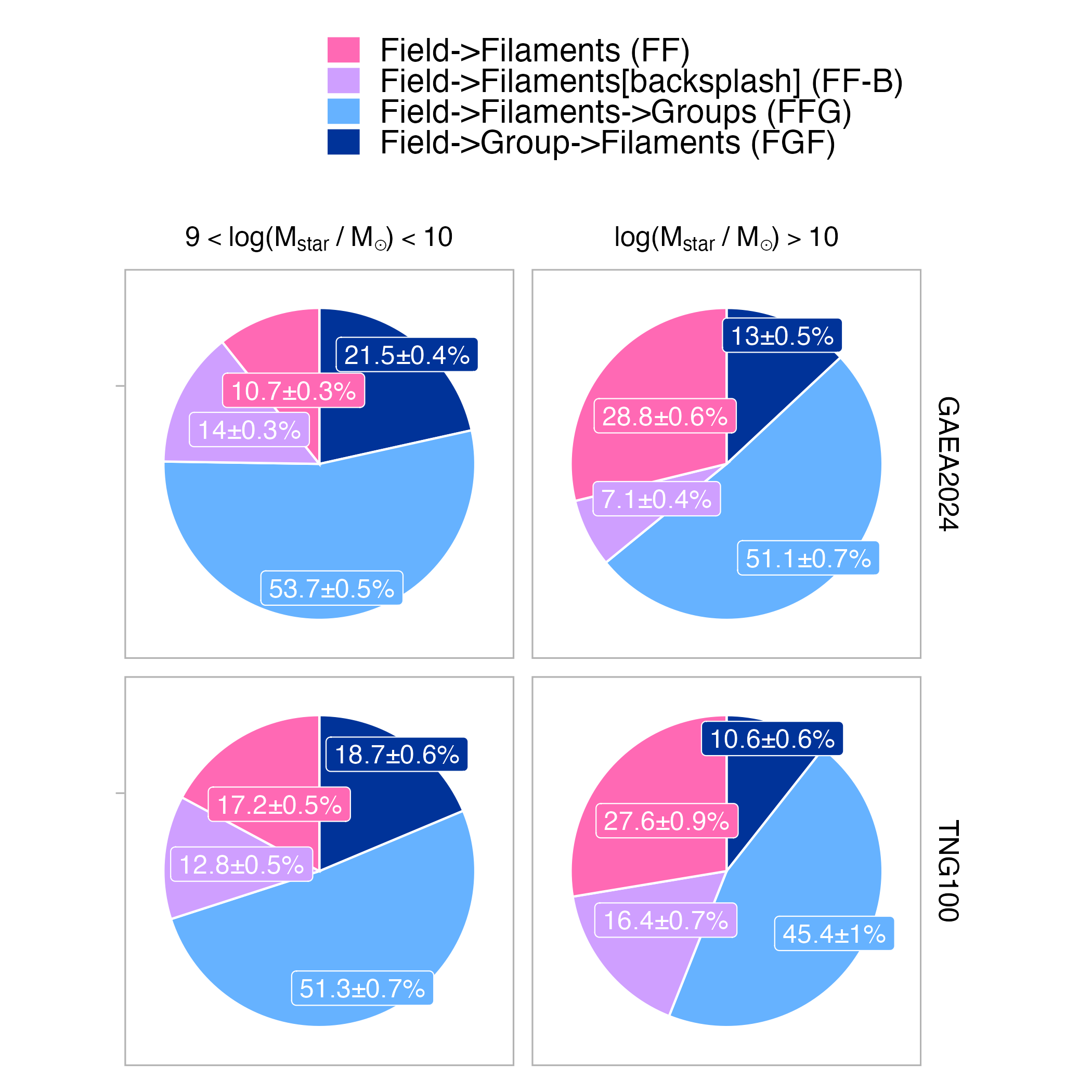}
    \caption{Pie charts showing the fraction of galaxies inside filaments 
    at z=0 separated by their environmental history and mass for GAEA (top) and TNG100 (bottom).
    The fraction $p$ and standard error~($SE = \sqrt{\frac{p \cdot (1 - p)}{n}}$) are indicated in each segment.}
    \label{fig:fils_massive_env_history_d1}
\end{figure}


\begin{figure*}
    \centering
   
    \includegraphics[width=1\linewidth]{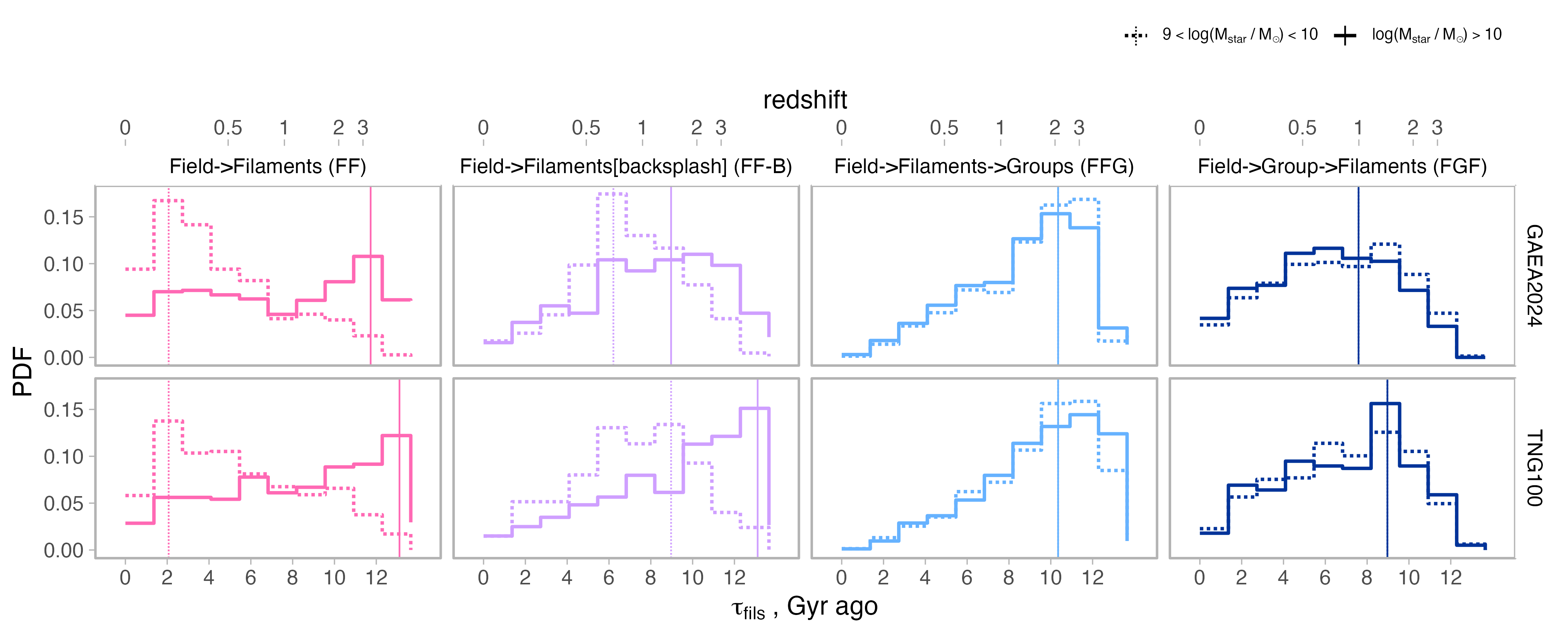}
    \caption{The filament infall times of $z = 0$ filament members are shown by the distribution of infall times ($\tau_{\rm fils}$) separated by environmental history and mass at $z = 0$ for two models. The vertical lines represent the peak values. }
    \label{fig:fils_massive_env_history_infalltimesfils}
\end{figure*}

Figure~\ref{fig:fils_massive_env_history_d1} shows the percentage of filament members that underwent the various environmental histories described in the previous section, for two stellar mass bins. 
Regardless of the stellar mass and model adopted, the majority of filament galaxies belong to the group environment for some fraction of their lifetime~(FGF, FFG and FF-B). Galaxies that are inside filaments and avoided group pre-processing (FF) represent less than 20\% of the low-mass galaxy population and ~30\% of the massive population. This figure highlights the significance of the role that the group environment plays in shaping the observed properties of filament galaxies. Indeed, around 50\% of the total filament population first fall into filaments and afterwards infall onto a group~(FFG). The fraction of galaxies experiencing this pathway is largest among the low-mass 
galaxies in GAEA (53.7$\pm$0.5\%), and lowest among the massive 
galaxies in TNG100 (45.4$\pm$1\%). The predominance of the other pathways depends on both the model and the stellar mass. At the massive end, the second most common pathway is FF (about 30\% for both models considered). The third most common path for massive galaxies in TNG100 is FF-B with 16.4$\pm$0.7\%, while GAEA2024 predicts only 7.1$\pm$0.4\% of galaxies in this environmental history channel. Overall, we find a larger incidence of backsplash galaxies in the hydrodynamical model for massive galaxies. At low masses, in the GAEA2024 model, the second most common path is FGF with 21.5$\pm$0.4\%, followed by FF-B~(14.4$\pm$0.3), and FF~(10.7$\pm$0.3\%). The corresponding fractions for  TNG100 are: FF with 18.7$\pm$0.6\%, then FF-B is 14.3$\pm$0.5\%, and FGF has 19$\pm$0.6\%. This means that TNG100 predicts a fraction of low-mass galaxies that were never processed as satellites, larger than in GAEA2024.
\par
We investigate how often galaxies outside the filaments\footnote{These galaxies do not enter the statistics shown in Fig~\ref{fig:fils_massive_env_history_d1}} at z=0 interacted with groups during their evolution. We stress that, in our definitions, galaxies outside filaments are not equivalent to field galaxies: while field galaxies are always outside filaments and have never been part of a group as satellites, many galaxies outside filaments may have experienced group interactions.
We find that a larger fraction of galaxies outside filaments than inside filaments have never interacted with groups: GAEA2024 predicts that 78.9$\pm$0.3\%~(88.4$\pm$0.4\%) of the low (high)-mass galaxies outside filaments have never been identified as satellites of any group. Considering TNG100, we get that 72.4$\pm$0.5\ (76.5 $\pm$1\%) of low (high)-mass galaxies never interacted with groups. Comparing these values with those corresponding to galaxies inside filaments, we conclude that it is much more likely for galaxies outside filaments than inside filaments to avoid any interactions with groups. This finding supports the idea that filaments act as bridges connecting groups and clusters \citep[e.g.][]{Tempel+2014_perls}, and that there is a higher probability that galaxies within filaments have experienced group environments at some point in their history.
\par
In addition to examining the distribution of filament galaxies across different environmental histories, it is crucial to quantify the timing of when galaxies became filament members, as this allows us to assess when they started to experience the influence of the filament and for how long they have been in such an environment. Figure~\ref{fig:fils_massive_env_history_infalltimesfils} shows the distribution of $\tau_{\rm fils}$  for galaxies corresponding to the different environmental paths described above. The two models give consistent overall results. Low-mass FF galaxies, i.e. galaxies that fell directly onto the filaments, are the ones with the youngest $\tau_{\rm fils}$ values (the peak is around 1.4 Gyr ago for both models). This is also the pathway where low and high-mass galaxies differ the most: distributions are skewed towards earlier epochs for more massive galaxies, with the corresponding peak of $\tau_{\rm fils}$ at $\sim 11-12.5$ Gyr in both models. Differences between low and high-mass galaxies are also clear in the FF-B case, even though they are less pronounced: in both models, low-mass galaxies spent less time in the filaments than their larger mass counterparts. For the two remaining channels, FFG and FGF, no clear differences are evident between the two mass bins. In the FFG case, the infall time distributions for both low-mass and massive galaxies are skewed towards early cosmic epochs with a peak at $\tau_{\rm fils}\sim 7-8$~Gyr ago, while the distributions for the FGF case peak around 9 Gyr ago. 
\par
In the latter two cases, we can measure the time galaxies spend in a filament before falling onto a group, as the difference between $\tau_{\rm fils}$ and $\tau_{\rm group~infall}$.  
According to GAEA, low (high)-mass filament galaxies spend on average 1.0$\pm$0.1 (1.1$\pm$0.2) Gyr before entering the group environment, and 75\% of all the galaxies spend less than 3 Gyr in filaments before falling into a group. These numbers rise to 1.7$\pm$0.2~(1.2$\pm$0.3) Gyr in TNG100, where 75\% of the population spend less than 3.5 Gyr in filaments before falling onto a group. 
If we instead consider the time galaxies spend in groups before infalling into filaments, we find 
an average duration of 1.5$\pm$0.3 Gyr in GAEA and 2.0$\pm$0.2 Gyr in TNG100, consistent across both low-mass and massive galaxies, with 75\% of the galaxies spending less than 3-4 Gyr as satellites of groups outside filaments in TNG100 (GAEA2024).
\par
We note that the fractions discussed above have a clear dependence on the definition of filaments, and especially on the radius chosen to define filament members. We discuss how results vary when considering different membership criteria for filaments in Appendix~\ref{appsec:env_hist_var}. Overall, regardless of the adopted filament radius, we find that more than half of the filament galaxy population is processed in groups, and the FF population is always a minority. 

\section{Dependence of z=0 filament galaxy properties on their environmental histories 
}
\label{sec:res_props}

In this section, we investigate how the properties of filament galaxies are shaped by their environmental history.  We begin by examining the stellar mass distributions of z=0 filament members, and tracking their stellar mass assembly over time, to assess whether the two models predict any significant influence of filaments and groups - such as halting or accelerating their stellar mass growth. We then explore the star-formation history and gas content of filament galaxies as a function of cosmic time, with a focus on how environmental factors like tidal fields or gas accretion through filaments may modulate star formation. 

\subsection{Stellar mass at z=0}
\label{subsec:stm_z0}

We begin by examining the stellar mass of galaxies residing in filaments. Previous studies have established that galaxies located within filaments tend to have higher stellar masses compared to their counterparts in the field \citep{Laigle+2018, Kraljic+2018, Singh+2020}. In this section, we investigate if the models considered reproduce these observed trends and to what extent these can be explained by a different environmental history. 

\par
\begin{figure}
    \centering
    \includegraphics[width=0.9\linewidth]{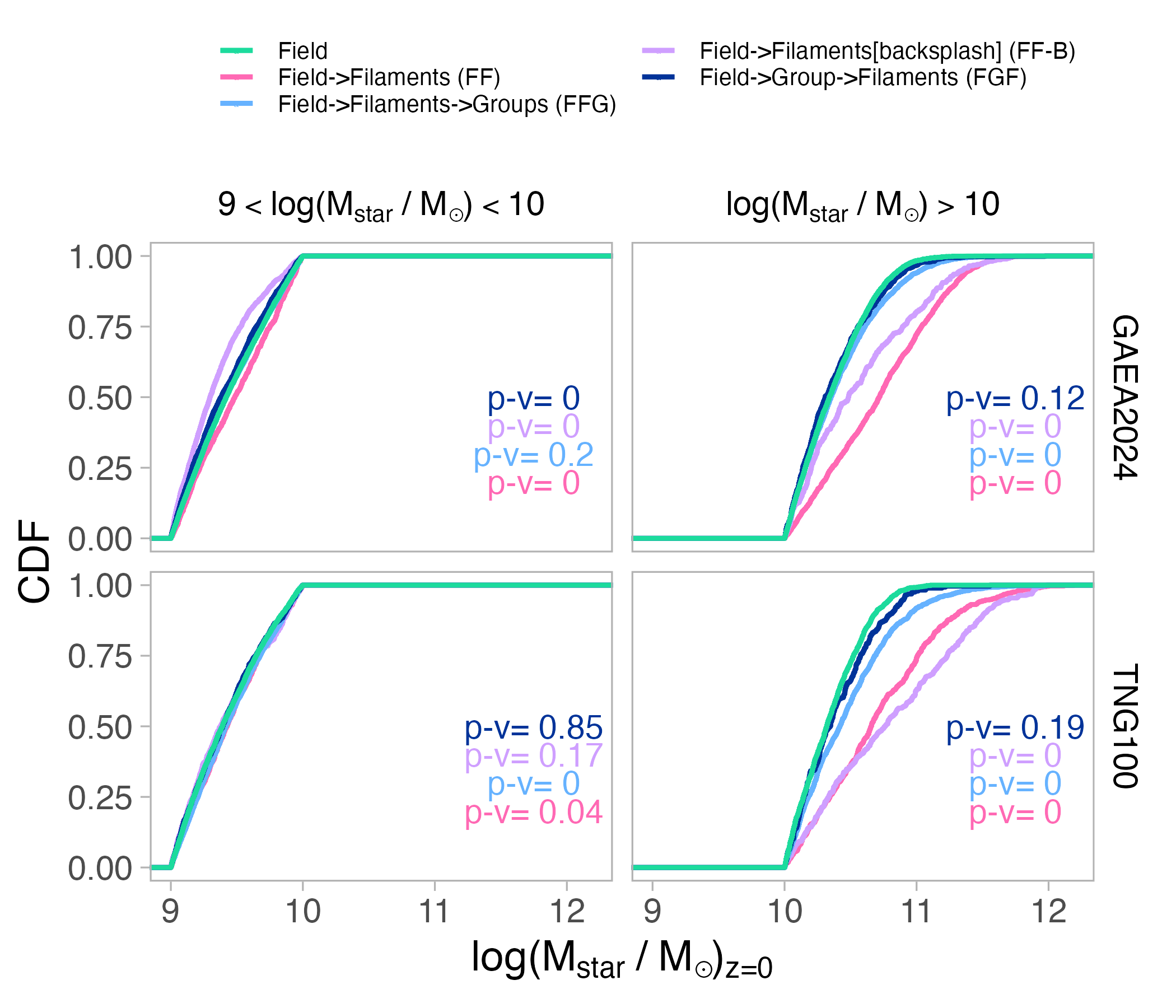}
    \caption{The cumulative mass distribution of stellar mass of galaxies inside filaments at z=0 separated by environmental history.  For comparison, the values for the field galaxies are also reported. The Kolmogorov-Smirnov test values for comparing filament and field galaxies are shown in each panel. }
    \label{fig:fils_stellarmass_byenvhist_cdf}
\end{figure}

Fig.~\ref{fig:fils_stellarmass_byenvhist_cdf} shows the cumulative stellar mass distribution of galaxies at z=0 residing in filaments, separated by their environmental history, alongside the distribution for field galaxies. The figure is complemented by Kolmogorov–Smirnov (KS) test results, which compare the stellar mass distributions of each filament population to those of the field galaxies. The test robustly indicates that massive galaxies exhibit a strong dependence of their z=0 stellar mass distribution on their environmental history, whereas low-mass galaxies show a much smaller dependence. For the low-mass populations in both models, the stellar mass distributions appear very similar to those of the field upon visual inspection of galaxies, although the KS still finds a significant difference between the stellar mass distributions of low-mass galaxies in filaments and in the field.
\par
The stellar mass distribution of massive galaxies depends significantly on their environmental history, with the FF and FF-B populations in both models showing a clear shift towards higher stellar masses. These two channels include the most massive galaxies, with 16\% and 4\% (12\% and 15.2\%) of galaxies exceeding $\log \rm (M_{star}/\rm M_{sun}) = 11$ in GAEA2024 (TNG100), respectively. The other environmental history channels include fewer than 2.5\% of galaxies above this mass threshold. For comparison, field galaxies include only 0.5\% and 0.2\% of galaxies above the same mass cut. Therefore, both models predict more massive galaxies within filaments than in the field, and a larger fraction of these galaxies were not preprocessed as satellites (i.e., FF or FF-B). Massive filament galaxies that were preprocessed as satellites tend to have lower stellar masses than the FF and FF-B populations, consistent with previous findings that group preprocessing quenches star formation and limits stellar mass growth \citep{De_Lucia+2012}.


\par
\begin{figure}
    \centering
    \includegraphics[width=0.9\linewidth]{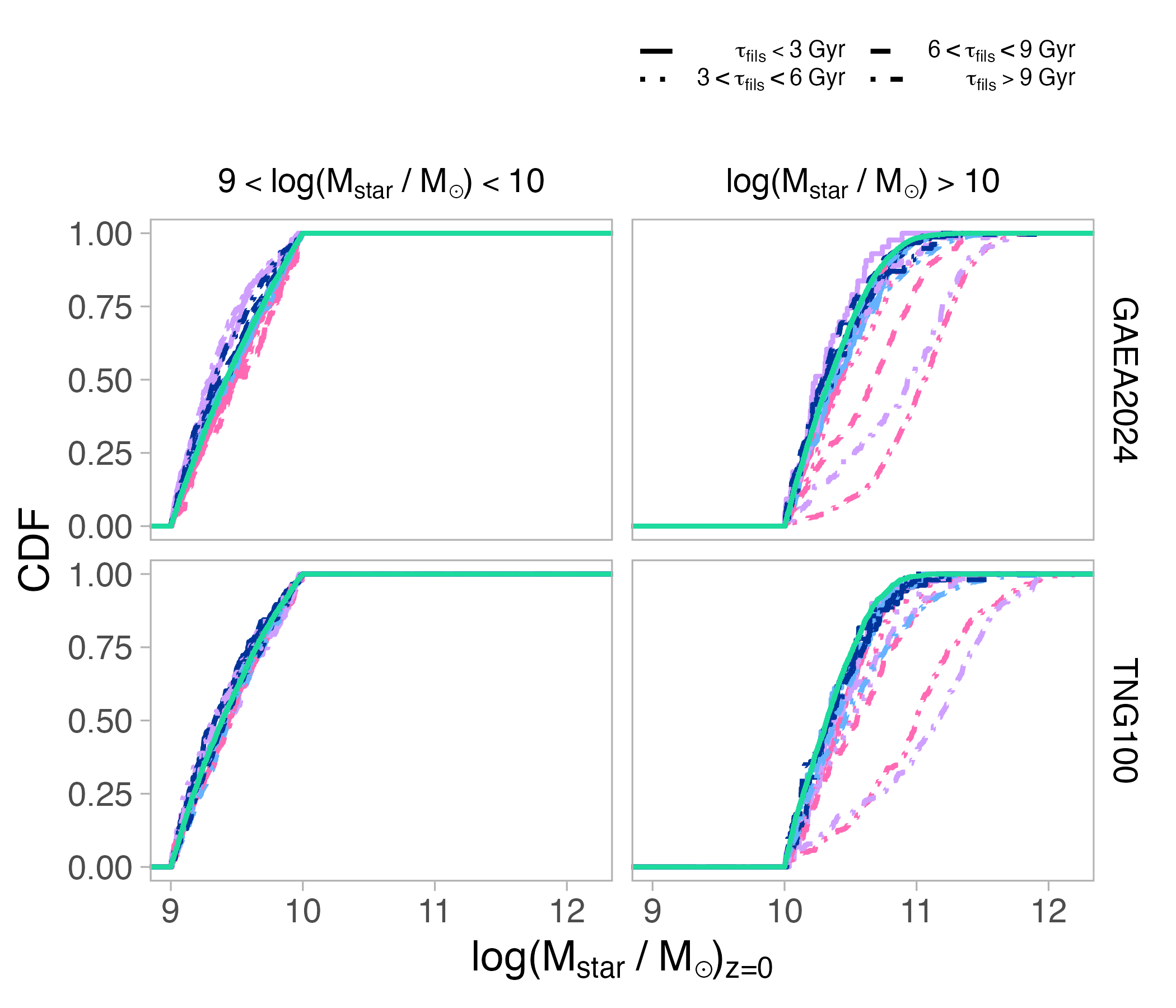}
    \caption{Same as Fig.~\ref{fig:fils_stellarmass_byenvhist_cdf}, but further split into bins of infall time into filaments.}
    \label{fig:fils_stellarmass_byenvhist}
\end{figure}

Furthermore, the stellar masses of the FF and FF-B populations depend on their infall times onto filaments, as shown in Fig.~\ref{fig:fils_stellarmass_byenvhist}. According to both GAEA2024 and TNG100, massive FF and FF-B galaxies that fell into filaments more than 9 Gyr ago shifted towards the massive 
end with respect to other populations of filament galaxies. For instance, FF galaxies accreted onto filaments less than 3 Gyr ago have a median stellar mass of $\log \rm (M_{star}/\rm M_{sun}) = 10.5\pm0.4$~($10.4\pm0.3$), while those accreted more than 9 Gyr ago have median masses of  $\log \rm (M_{star}/\rm M_{sun}) = 11.1\pm0.3$~($11.0\pm0.4$). No such dependence on infall time is found for low-mass galaxies or other environmental history channels for massive galaxies. This suggests that the filaments favour mass growth~(if galaxies are not satellites), and the shift towards more massive filaments galaxies reported in previous works is due to FF and FF-B populations. 


\subsection{Stellar mass assembly history}
\label{subsec:stm_assembly}

We now focus on how galaxies accumulated their stellar mass over time, depending on their environmental history. To minimise biases introduced by differences in stellar mass distributions, we consider mass-matched samples.  We consider the evolution of the large-scale structure and filaments, as well as the time spent by galaxies within filaments. 
We randomly select galaxies at z=0 from the four environmental history channels considered, divided in bins based on their $\tau_{\rm fils}$: $\tau_{\rm fils}$< 3 Gyr ago,  3$<\tau_{\rm fils}<$ 6 Gyr ago, 6 $<\tau_{\rm fils}<$ 9 Gyr ago, and $\tau_{\rm fils}$> 9 Gyr ago. We also select a control sample of field galaxies. For each of these populations and in any given time bin, we match the stellar mass distribution to that of the field sample. 
We repeat the random selection 100 times. For each realisation of the stellar mass-matched sample, we compute the median and 1$\sigma$ confidence interval (via bootstrapping) of the following measured quantities: stellar mass fraction, quenched fraction, and gas content. We then report the overall median as the median across all realisations and define the uncertainty range using the minimum~(maximum) of the lower~(upper) bounds from the 100 realisations.

\begin{figure*}
    \centering
    \includegraphics[width=1\linewidth]{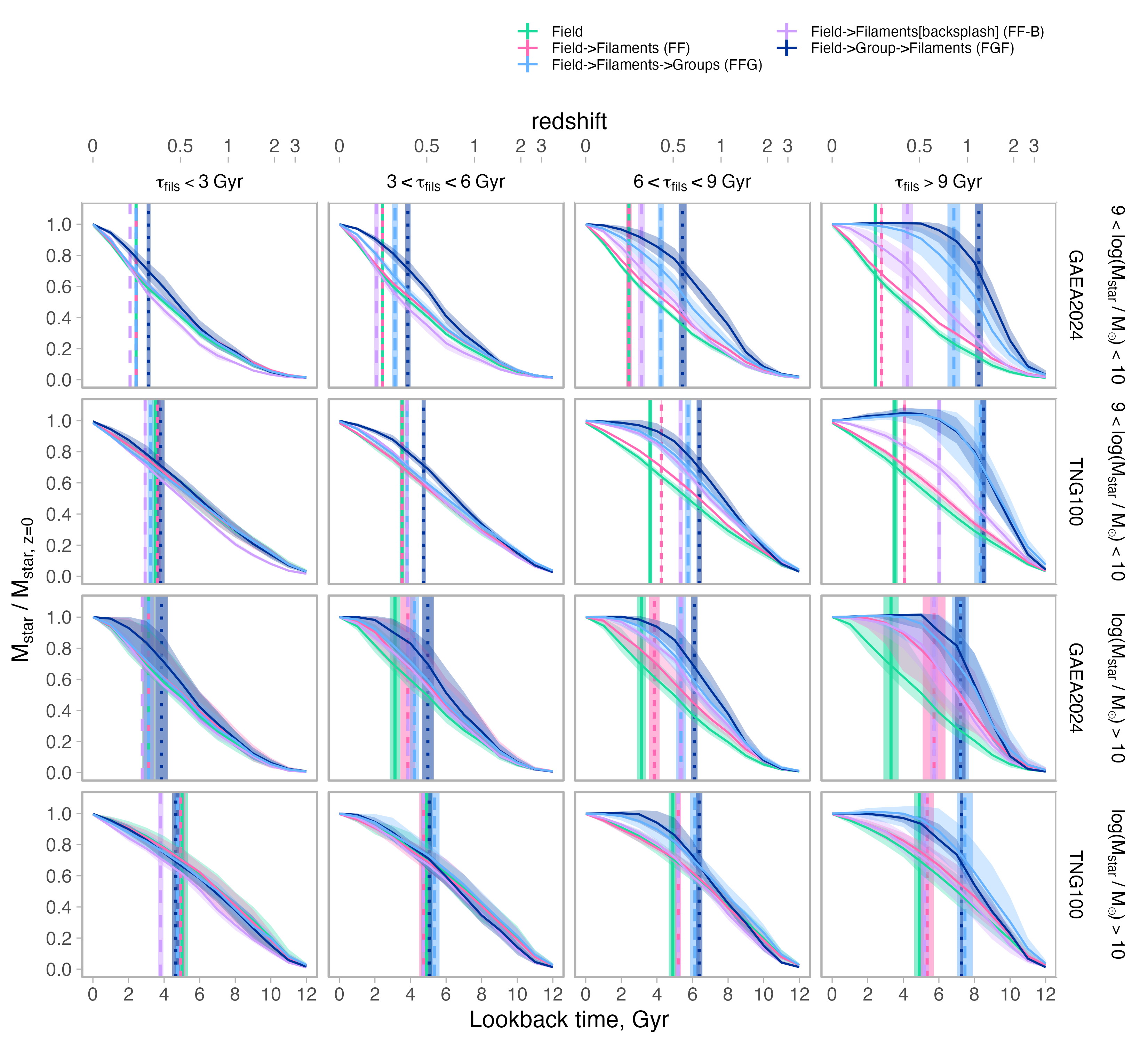}
    \caption{Median cumulative fraction of stellar mass $M_{\rm star}/M_{\rm star, z=0}$ for today's low-mass~(first and second rows) and massive~(third and fourth rows) members of filaments categorized by environmental history and infall time into filaments, compared to field galaxies~(stellar mass is controlled in each panel). Results are shown for two models with solid lines representing the median and shaded regions indicating uncertainty limits across 100 stellar mass-matched realisations. The vertical lines represent the median time when galaxies build 70\% of their mass, with a standard deviation between 100 stellar mass fittings.}
    \label{fig:sfh_fils_by_branch_time}
\end{figure*}


\par
Fig.~\ref{fig:sfh_fils_by_branch_time} presents the stellar mass assembly history for filament galaxies that experienced different environmental pathways. The stellar mass growth is traced by the mass of the main progenitor at each snapshot, normalised by the z=0 mass: $M_{\rm star}/M_{\rm star, z=0}$. To test the robustness of the observed trends, we repeated the analysis after randomly shuffling the environmental history labels, following an approach similar to that of \cite{Kraljic+2018}. In this case, the differences between the environmental branches are no longer apparent.
This confirms that the results are statistically robust and accurately reflect the impact of environmental history on the assembly of stellar mass in galaxies.
\par
The figure shows that the impact of environmental history on stellar mass assembly becomes stronger with earlier infall times onto filaments, for both low-mass and massive galaxies, in both models. Indeed, the stellar mass assembly histories of galaxies that fell onto filaments less than 3 Gyr ago~(leftmost column) are close to those of the field galaxies, regardless of the environmental history or stellar mass. In contrast, for galaxies with  $\tau_{\rm fils}$ > 9 Gyr ago (rightmost column) the median stellar mass assembly histories vary significantly across different environmental channels,  both for low-mass and high-mass galaxies, in both models. More specifically, galaxies that fell onto filaments less than 3 Gyr ago assembled 70\% of the z=0 stellar mass around 3-4 Gyr ago, largely independent of their environmental histories. For the FF and FFG channels, it is expected that such galaxies spend the majority of their life as field members, and  3 Gyr may be too short a time for filaments to affect the stellar mass assembly. This suggests that the influence of the filament and group environment on stellar mass assembly is not instantaneous and requires a prolonged residence within filaments to become evident.
\par
We now focus exclusively on the rightmost panels of Fig.~\ref{fig:sfh_fils_by_branch_time}, which show the most prominent trends. For low-mass galaxies, 
(top two rows) distinct stellar mass assembly histories for all populations of filament galaxies compared to field galaxies are evident, most notably for the FGF, FFG, and FF-B populations, which have experienced a group environment. Specifically, low-mass FGF and FFG  galaxies assembled their stellar mass earlier, reaching 70\% of their present-day mass approximately 6–8 Gyr ago, followed by a clear truncation of stellar growth around 3–5 Gyr ago. In contrast, field galaxies reached 70\% of their present mass much later (around 2.4$\pm$0.1 Gyr ago in GAEA2024 and 3.5$\pm$0.2 Gyr ago in TNG100) and continued to build up their stellar mass down to the present. This earlier assembly and subsequent truncation are seen only in low-mass filament galaxies that were processed as satellites, suggesting that interactions with halos strongly suppress further stellar mass growth. This is consistent with the well-known efficient quenching of group environments, especially for low-mass galaxies~\citep{De_Lucia+2012, Xie+2020, Donnari+2021_qf}.
Focusing on the FF population, which has never experienced group processing,
we find that they continue to build up stellar mass over time, and they assembled 70\% of their stellar mass 2.7$\pm$0.1 Gyr ago and 4.1$\pm$0.1 Gyr for GAEA2024 and TNG100, which is slightly shifted toward earlier assembly compared to the field. We therefore conclude that filaments only slightly enhance stellar mass assembly for low-mass galaxies, if they have not undergone satellite processing.  The absence of any inflection in the growth rate of stellar mass suggests that the tidal field of the cosmic web does not significantly hinder stellar mass growth in low-mass galaxies. Instead, group processing, including that occurring for the backsplash population, appears to be the dominant factor distinguishing galaxies in filaments from their field counterparts. 
\par
Moving on to massive galaxies (bottom two rows in the rightmost panels of Fig. 8), we find a broadly similar picture: their stellar mass assembly histories differ significantly across different environmental history channels of filament galaxies compared to the field. As for low-mass systems, the satellite population of FGF and FFG shows the most pronounced divergence from the field. However, even massive FF galaxies exhibit distinct assembly histories, highlighting the role of filaments in shaping the assembly history of massive galaxies. This trend is particularly pronounced in  GAEA2024, which exhibits a substantial shift in stellar mass assembly for the FF population compared to the field. Massive FF galaxies in this model reached 70\% of their present-day stellar mass around 5.5$\pm$0.2 Gyr ago, and have experienced minimal stellar mass growth over the past 3 Gyr. In contrast, massive field galaxies are still actively assembling their stellar mass, having reached the 70\% assembly threshold only around 3.3$\pm$0.2 Gyr ago. The TNG100 model also predicts a faster assembly for FF galaxies relative to the field, though the effect is less pronounced and does not show a clear reduction of the stellar mass growth rate at recent cosmic times. In TNG100, FF galaxies reached 70\% of their mass by 5.5$\pm$0.2 Gyr ago, compared to 4.9$\pm$0.1 Gyr ago for field galaxies.

\subsection{Star-formation of filament populations}
\label{subsec:sfr}

\begin{figure*}
    \centering
    \includegraphics[width=1\linewidth]{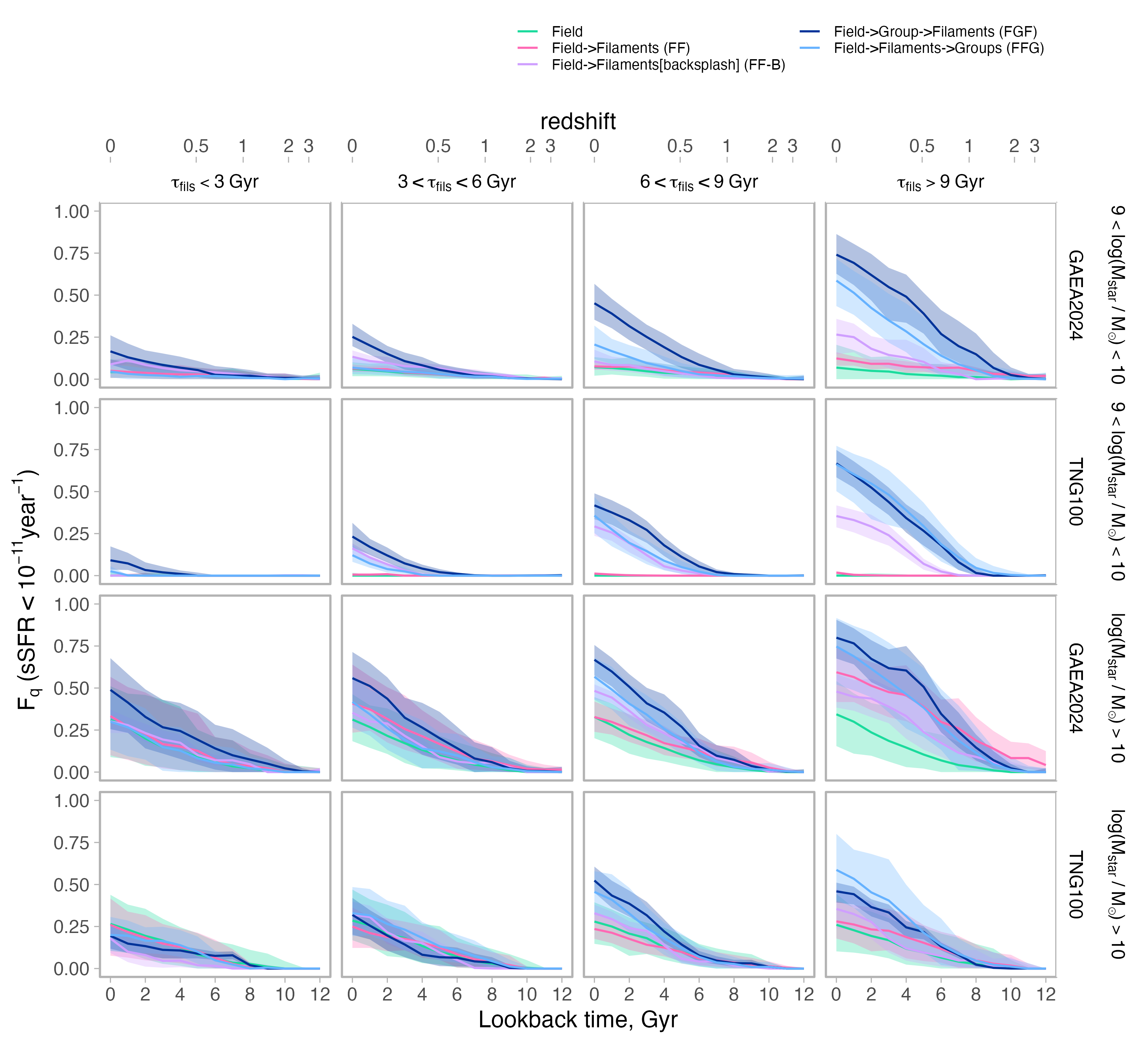}
    \caption{Quenching of the filament populations over time compared to the field population. Quenched fraction of filament population as a function of cosmic time, separated by the environmental history and infall times into filaments~(stellar mass is controlled). The galaxy is considered quenched if $sSFR$ is below $10^{-11}$ year$^{-11}$. Results are shown for two models with solid lines representing the median and shaded regions indicating uncertainty limits across 100 stellar mass-matched realisations. We note that for the GAEA2024 model, the star-formation rate was time-averaged, and for TNG100, it was instantaneous. Therefore, comparisons between the models at this stage should be interpreted qualitatively.  }
    \label{fig:sfr_fils_by_branch_time}
\end{figure*}


In this section, we discuss how different populations of filament galaxies quench over time, in comparison to the field. Figure~\ref{fig:sfr_fils_by_branch_time} shows the evolution of the quenched fraction for galaxies within filaments as a function of cosmic time, for different environmental history channels and infall time onto filaments.

Again, the difference in quenching fractions between filament and field galaxies, as well as among different filament populations, strongly depends on $\tau_{\rm fils}$. In the leftmost panel (corresponding to $\tau_{\rm fils} < 3$ Gyr ago), the differences are small or negligible in both models and across both stellar mass bins. However, galaxies that accreted onto filaments earlier exhibit more pronounced differences in their quenching fractions compared to the field. This confirms that environmental history influences the star formation history of galaxies. Again, we discuss the rightmost column since it shows the most pronounced trend. 
\par
Focusing on low-mass galaxies, both models predict that the most quenched populations are satellite filament galaxies from the FGF and FFG: 0.74$\pm$0.04 (0.65$\pm$0.02) and 0.58$\pm$0.04 (0.67$\pm$0.02) of these galaxies are quenched by z = 0 in GAEA2024 (TNG100), and their quiescent fractions exceed those of field galaxies for at least the last 8 Gyr of evolution. The third most quenched population is FF-B, which also consists of galaxies that interact with halos as satellites $\rm F_q = 0.25-0.35$. In contrast, the FF population of low-mass galaxies exhibits a slightly elevated quiescent fraction compared to the field only in GAEA2024, reaching 0.12$\pm$0.02 at z = 0, while the field reaches 0.08$\pm$0.03. TNG100 predict no low-mass quenched galaxies in the field, and around 0.01 for the FF population. These results suggest that, for low-mass filament galaxies, quenching is primarily driven by group processing, with minimal contribution from the filamentary environment itself, even in TNG100.
\par
For massive galaxies, we also find a distinct $\rm F_{q}$ evolution over time between field and filament populations. Again, the most quiescent populations are filament satellites from FGF and FFG: 0.76$\pm$0.04 or 0.64$\pm$0.03 and 0.74$\pm$0.05 or 0.58$\pm$0.04 up to z = 0, for GAEA2024 and TNG100, respectively. The FF-B and FF populations have quenched fractions of 0.47$\pm$0.06 and 0.58$\pm$0.05 in GAEA2024, and 0.35$\pm$0.03 and 0.28$\pm$0.03 in TNG100. In contrast, the field values are relatively lower, 0.33$\pm$0.05 or 0.26$\pm$0.02. Therefore, all filament populations have a higher fraction of quiescent galaxies, even though they have never been processed as satellites~(FF). This effect is especially prominent in GAEA2024 and,  similarly to the stellar mass assembly, we attribute it to different assembly histories of galaxies inside filaments and the field. These differences could be driven by higher merger rates in dense environments~\citep{Rodriguez-Gomez+2015} or stronger AGN feedback~\citep{Weinberger+2018}. In TNG100 at z=0, the differences are not as pronounced as in GAEA2024, but the quenching pathway of the FF population differs significantly from that of field galaxies at any cosmic time. 
\par
We verified that our results are robust against different choices of quenching criteria. In particular, adopting a threshold of more than 1 dex below the star-forming main sequence in the SFR–$M_{\star}$ plane yields consistent trends.

\section{Discussion}
\label{sec:discussion}
We firstly note that we find no strong evidence that galaxies are significantly affected by the cosmic web tidal field in the TNG100 model: (i) stellar mass growth in the FF population is not suppressed in either model in any mass bin; (ii) in TNG100, low-mass FF galaxies do not show enhanced quenching relative to the field. While a tidal field impact cannot be fully excluded, its effect appears negligible for the FF population studied here and may only become relevant for galaxies below our mass threshold ($\rm \log(M_{star}/M_{sun}) < 9$).
\par
In the previous Section, we showed that models predict little direct impact of filaments on the stellar mass distribution and quenched fractions for low-mass galaxies. In contrast, group interactions clearly leave a strong imprint: low-mass galaxies in the FGF and FFG populations exhibit larger quenching fractions than filament galaxies that did not experience the group environment. Nevertheless, we still observe a systematic offset in stellar mass assembly histories for FF galaxies compared to the field. This may reflect a higher availability of material for accretion inside the cosmic web, potentially linked to enhanced merger rates or differences in the stellar mass acquired through mergers. 
Besides, GAEA2024 predict a very prominent difference between the field and the FF population. At the same time, this mode does not implement any filament-specific physical prescription; this effect is likely driven solely by differences in merger trees of haloes residing in different environments. We explore the role of mergers in shaping the results of the previous section here.

\subsection{Merger history as explanation of different stellar mass assembly}

In Sections~\ref{subsec:stm_z0} and \ref{subsec:stm_assembly}, we demonstrated that filaments promote the accumulation of stellar mass in massive galaxies compared to field 
and group counterparts. Naturally, filament galaxies that are matched to the field population by stellar mass distribution may have already had more massive main progenitors at $z=4$, implying that environmental effects could have begun influencing them even before the epoch considered here. However, by z=0, their stellar masses converge with those of field galaxies — although they follow distinct evolutionary pathways. In particular, the FF population shows a faster stellar mass assembly and a higher fraction of quenched galaxies compared to their field counterparts at fixed stellar mass. We propose three potential explanations for this finding: (1) a higher merger fraction among filament galaxies; (2) larger stellar masses of mergers; and/or (3) higher star-formation rates due to an excess of cold gas.
In the framework of the models studied, there is no enhancement of star formation within filaments. In fact, we find the opposite: the fraction of quenched galaxies is larger for the FF population. We therefore focus only on the mergers of field and FF galaxies. We examine the merger histories of galaxies as a function of their environmental evolution. Specifically, we extract all progenitors of each galaxy at 
z=0 and identify mergers by grouping galaxies with the same descendant. We classify merger events as minor when the stellar mass ratio $\mu = \rm M_{star,1} / \rm M_{star,2}$ lies between 1/10 and 1/4, and as major when $\mu  > 1/4 $~(only mergers with $\rm \log (M_{star} / M_{sun})$ > 7 are considered).  For each galaxy and at each snapshot, we count the number of mergers of each type. When more than one merger of a given type occurs at the same snapshot, we compute the corresponding median stellar mass of the minor and major mergers.  From now on, we will discuss only results for filament galaxies that were accreted onto filaments more than 9 Gyr, as they exhibit the most pronounced differences with respect to field galaxies.
\par
\begin{figure}
    \centering
    \includegraphics[width=1\linewidth]{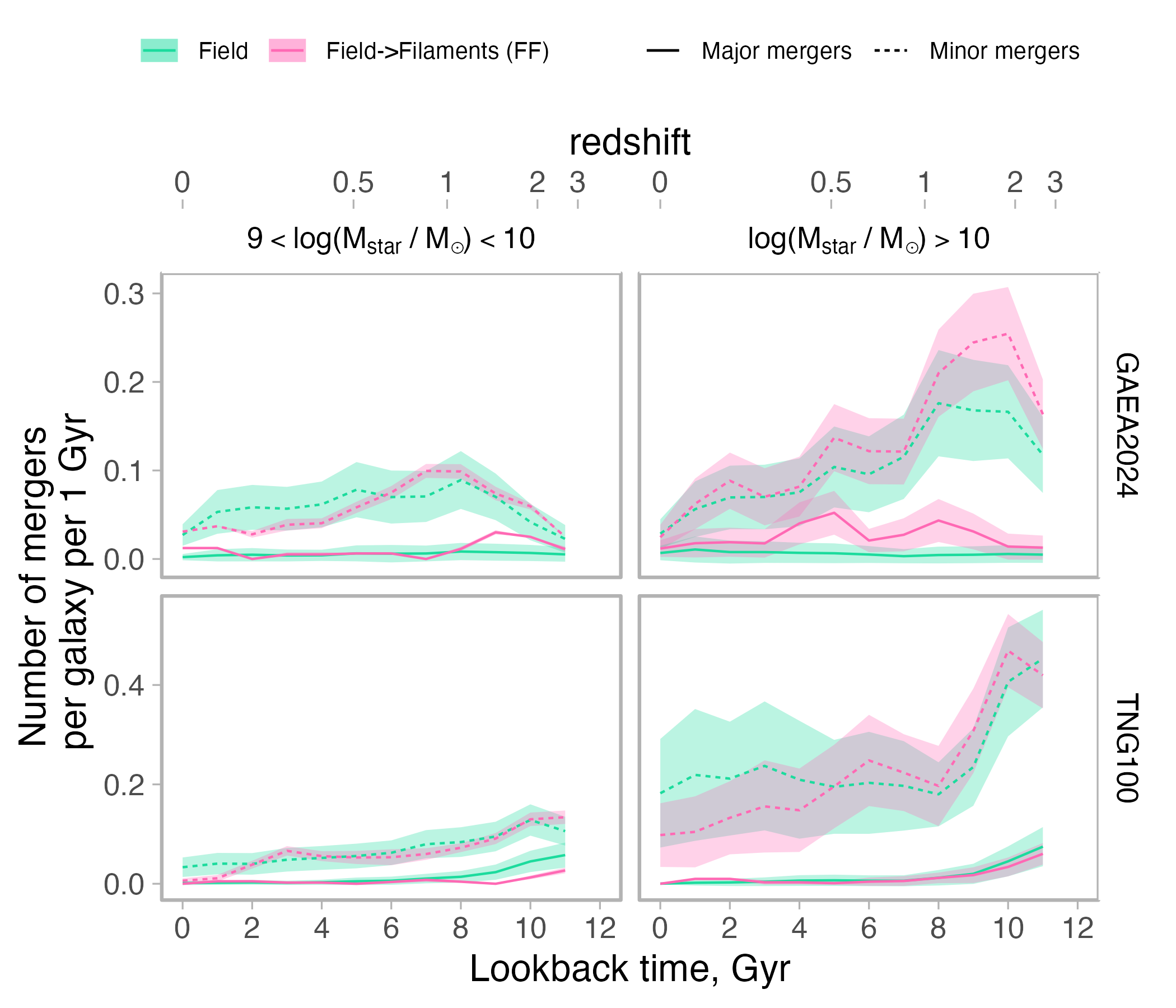}
    \caption{The merger rate for filament members fell into filaments more than 9 Gyr ago. The lines represent the median number of major~(solid line) and minor~(dotted line) mergers per galaxy per 1 Gyr as a function of cosmic time, categorised by environmental history when stellar mass is controlled. The values represent the total number of major mergers normalised by the total number of galaxies considered, with the figure showing the median across 100 realisations of stellar mass matching, and the shaded area demonstrates a standard deviation. }
    \label{fig:number_of_mergers_major}
\end{figure}

Fig.~\ref{fig:number_of_mergers_major} presents the number of minor and major mergers experienced by one galaxy per 1 Gyr over cosmic time for FF galaxies compared to the field population,  when stellar mass is controlled for. For low-mass galaxies, we do not observe an enhanced merger rate~(neither major nor minor), compared to the field in either model considered. However, in GAEA2024, the median total stellar mass acquired throught mergers over all time is slightly larger for the FF population~($\rm \log M_{star, total} / M_{sun} = 8.0 \pm  0.1$) than for field galaxies ~($\rm \log M_{star, total} / M_{sun} = 7.8 \pm  0.05$). In contrast, TNG100 does not predict a significant difference in the total accreted mass between the FF and field populations, with both around 7.8$\pm$0.1. 
\par
For the massive FF population, we find the same minor merger rate in both models, with scatter between realisations, but GAEA2024 also predicts an elevated major merger rate. The higher major merger rate measured for the FF population explains the significant difference in stellar mass assembly history between the FF and field populations in GAEA2024 (see Figure~\ref{fig:sfh_fils_by_branch_time}). The absence of such an enhancement in  TNG100  is consistent with the smaller difference in stellar mass assembly found between the two populations for this model. Moreover, the total stellar mass of merged galaxies in GAEA2024 of FF population reaches $\rm \log M_{star, total} / M_{sun} = 9.2 \pm  0.2$ which is larger than that found for the field  ($\rm \log M_{star, total} / M_{sun} = 8.6 \pm  0.1$). In TNG100, we do not find a significant difference, and the total stellar mass of FF and field mergers is comparable  $\rm \log M_{star, total} / M_{sun} = 8.6 \pm  0.2$.

\par
In conclusion, variations in major or minor merger rates cannot account for the stellar mass assembly differences of low-mass filament galaxies, as both models predict only minor variations across populations. For the FF population, GAEA2024 attributes the difference to more massive mergers, while TNG100 does not. Overall, however, the differences for low-mass galaxies, and in TNG100 in particular, remain small.
\par
For massive galaxies, the two models diverge. In GAEA2024, FF galaxies show a markedly different stellar mass assembly from the field (Figure~\ref{fig:sfh_fils_by_branch_time}), driven by higher rates of both major and minor mergers with more massive companions. In TNG100, merger rates are similar across populations, but the mergers that do occur involve more massive galaxies, producing a divergence in stellar mass assembly. These results suggest that galaxies accreted onto filaments earlier tend to merge with more massive neighbours, allowing them to build up larger stellar masses.
\par
In the Appendix~\ref{appsec:mergers_full}, we present the merger rates for all other bins of infall into filaments for both FF and field populations. Overall, the observed trend is weaker for galaxies that were accreted onto filaments in the last 9 Gyr, consistent with the smaller differences in stellar mass assembly reported earlier.
\par
We note that, in TNG100, the differences in stellar mass assembly largely disappear when controlling for stellar mass. However, this does not mean that filaments are irrelevant to the evolution of galaxies in this model. The most massive galaxies are still found exclusively within filaments and only when group processing is absent. These galaxies are typically central group members that grow through the accretion of satellites, which are more numerous in groups embedded in filaments~\citep{Guo+2015}. A similar but even stronger trend is found in GAEA2024.
\par
Finally, Dulcien et al. (2025, in prep.) report observational evidence for an enhanced galaxy merger rate in filaments, based on elevated galaxy-pair counts in cluster-connected filaments in the 4MOST CHANGES survey relative to both field and cluster environments, supporting our findings.

\subsection{Excess of quenched galaxies in FF population}

In Sec.~\ref{subsec:sfr}, we demonstrated an excess of quenched galaxies inside filaments when controlling for stellar mass. At z=0, massive galaxies that were accreted onto filaments more than 9 Gyr ago exhibit a higher quenched fraction than field galaxies in both models. This result is expected for galaxies that have interacted with groups as satellites, as groups are known to host a higher fraction of quenched populations~\citep{Wetzel+2012, Donnari+2021_qf}. However, our findings also reveal that this trend extends to the FF population, which has never interacted with haloes as satellites and thus has not experienced the environmental influence of groups. In this section, we discuss these findings in more detail, with a special focus on the role of filaments. 
\par
To better understand the nature of galaxy quenching inside filaments, we estimate the quenching timescales. Our approach is primarily based on the approach used in  \citet{Walters+2022}. In brief, we analysed the evolution of the specific star formation rate (sSFR) over time for the main progenitor of each galaxy identified at $z=0$. First, for each model, we determined how the sSFR evolves with lookback time for all star-forming galaxies at $z=0$, assuming they remained star-forming throughout their history. This establishes a reference "main sequence" of sSFR evolution for galaxies that have not experienced quenching. Quenching is then defined as a deviation below this sequence. We then identified the quenching starting time ($t_{\mathrm{start}}$) as the point when a galaxy's sSFR first falls below the $1\sigma$ lower bound of the main sequence. A quenching time ($t_{\mathrm{quench}}$) is defined as the moment when the galaxy's sSFR drops below $10^{-11}~\mathrm{yr}^{-1}$. The quenching timescale is then given by the time difference between $t_{\mathrm{start}}$ and $t_{\mathrm{quench}}$. Based on this timescale, galaxies are classified as: fast-quenching ($\leq 1~\mathrm{Gyr}$), slow-quenching ($\geq 2~\mathrm{Gyr}$), or rejuvenated (if their sSFR increases above the $10^{-11}~\mathrm{yr}^{-1}$ again after quenching, and this lasts for at least two subsequent snapshots). 
\par
\begin{figure}
    \centering
    \includegraphics[width=1\linewidth]{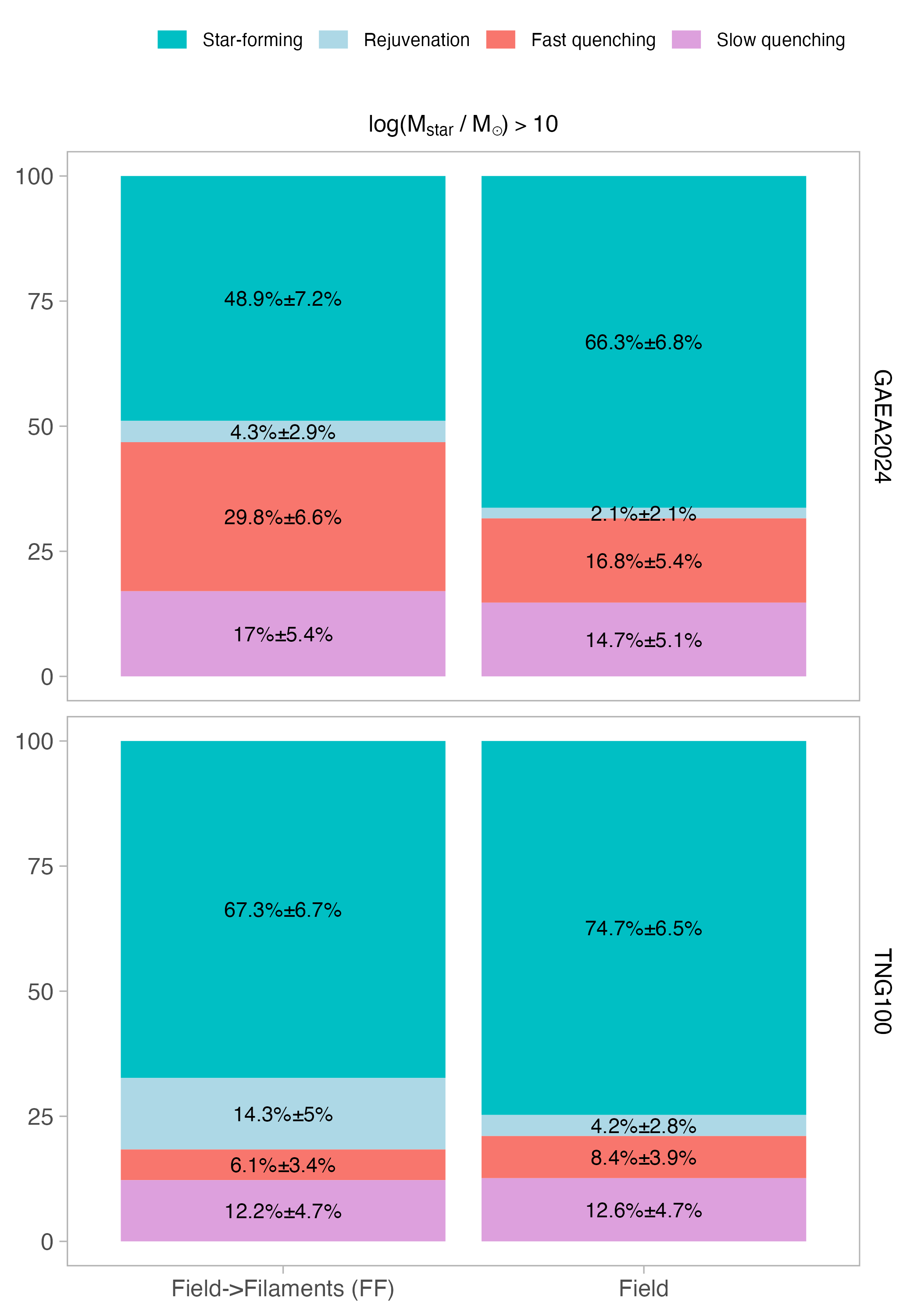}
    \caption{The distribution of the fraction of galaxy types by star formation or quenching status is presented for galaxies within filaments~(FF) and fields at z=0, with stellar mass controlled. Only massive galaxies that fell into filaments more than 9 Gyr ago are included. Populations are categorised into star-forming, rejuvenated, fast-quenched, and slow-quenched based on the evolution of their sSFR. The fractions were first calculated.}
    \label{fig:fils_fields_quenching}
\end{figure}

The results of this classification for massive FF galaxies at z = 0  and entering filaments more than 9 Gyr ago are shown in Fig.~\ref{fig:fils_fields_quenching} alongside field galaxies of the same stellar mass. 
\par
In GAEA2024, around 70\% of the field galaxies remain star-forming at z=0, while the quiescent population is evenly split between fast- and slow-quenching galaxies, each comprising approximately 15\%. For the FF population of massive galaxies, only 50\% remain star-forming at z=0, while the fractions of slow-quenching and rejuvenated cases are comparable to those in the field. Therefore, the excess of quenched galaxies in the FF population must be driven by fast quenching processes. Out of the ~30\% of fast-quenched FF galaxies, about half (i.e. ~15\%) could be attributed to internal processes similar to those responsible for fast quenching in the field population. This implies that the remaining ~15\% must be linked to environmental mechanisms specific to the FF population. Since GAEA2024 does not include stripping mechanisms within filaments, the most plausible fast quenching channels are AGN feedback and merger events. To test this, we examine whether merger events occur in GAEA2024 within 1 Gyr before the fast quenching episode. We find a notable difference between the FF and field populations: $56$ per cent of fast-quenched FF galaxies can be linked to mergers, whereas only 35 per cent of fast-quenched field galaxies have experienced mergers at the time of quenching (this value should be considered an upper limit). We also note that GAEA2024 predicts a negligible impact of rejuvenation, with similar rates between the field and FF populations
\par
The TNG100 model predicts that massive FF galaxies and field galaxies have the same fraction (around 70\%) of star-forming systems at z=0. Moreover, the fractions of fast- and slowly quenched massive galaxies are consistent within uncertainties between the FF and field populations. The only notable difference between the two environments is that TNG100 predicts a significantly higher fraction of rejuvenated galaxies inside filaments compared to the field. This implies that TNG100 primarily associates filaments with multiple episodes of quenching and rejuvenation, without specifically enhancing the fractions of fast or slow quenching. This highlights important differences between the models: TNG100 predicts more frequent rejuvenation events in filaments compared to GAEA2024, along with a lower fraction of rapidly quenched galaxies in these environments.

%

\section{Conclusions}
\label{sec:conclusion}

In this paper, we reconstruct the environmental history of filament galaxies at z=0 to better understand the role of environment (specifically filaments and the groups embedded within them) in galaxy evolution. Our approach introduces new metrics, such as the infall time into filaments and groups and their environmental histories, which have not been considered in previous studies. Indeed, disentangling the influence of filaments and of the groups embedded within them remains challenging. Our approach enables us to isolate the impact of different evolutionary pathways of filament galaxies on their present-day properties by comparing the Field->Filaments population with the control sample field and group population within filaments.
\par
Our analysis demonstrates that filaments host a heterogeneous population of galaxies with diverse environmental histories and infall times. Most galaxies within filaments have spent part of their evolution as satellites~(FFG, FGF and FF-B populations). This emphasises the importance of groups in shaping the statistical properties of filament galaxies, and at the same time highlights that filaments are preferential sites for finding groups~(FFG is $\sim$50\% of the filament population and found their host group in 1-3 Gyr after infalling into a filament). Only a minority, 10–20\% of low-mass galaxies and ~30\% of massive galaxies, have never been satellites~(FF population) and were influenced solely by the filamentary environment. 
\par
Despite significant differences in the numerical approach and assumptions for specific physical processes, GAEA2024 and TNG100 give consistent results for the environmental history of filament galaxies, with the notable exception of the backsplash population. The hydrodynamical model TNG100 predicts a higher fraction of massive backsplash galaxies than GAEA2024~(16\% and around 7\%, respectively).
\par
We also demonstrate that the environmental history of galaxies is affecting the stellar mass assembly and quenched fraction of the population. The impact of the environment on galaxies becomes increasingly significant with time, as they fall into filaments and groups. For example, 
we find a significant difference in stellar mass assembly and quenching fraction for those galaxies that fell onto a filament more than 9 Gyr ago, regardless of the mass bin or model, while galaxies that entered filaments more recently show little difference from their field counterparts. As a result, the mixing of galaxies with different environmental histories and infall times makes it challenging to isolate and detect the specific role of filaments in galaxy evolution in observational data, where only today's environment is available.
\par
In particular, our results show that low-mass galaxies ($ 9 < \log_{10}(\rm M_{\rm star}/\rm M_{sun}) < 10$) in filaments are primarily regulated by group processing, while the direct effect of the filamentary environment appears minimal. We find that galaxies reside inside filaments but have never been processed as satellites~(FF) evolve very similarly to the field population, with only a mild enhancement in stellar mass assembly and with no increase in quiescent fraction according to both models. In contrast, low-mass galaxies that experienced a group environment (FGF, FFG, FF-B) assembled their stellar mass earlier and show strongly elevated quenching fractions, consistent with gas removal processes that truncate star formation. These results align with previous work showing that low-mass galaxies are particularly sensitive to group-driven environmental effects~\citep{Xie+2020, Donnari+2021_qf}.
\par
Filaments play a decisive role in the evolution of massive galaxies  $\log_{10}(\rm M_{\rm star}/\rm M_{sun}) > 10$. Specifically, we find that galaxies which entered filaments and remained centrals show a clear dependence of their z=0 stellar mass on the time spent inside filaments. Filaments provide larger access to material for stellar mass assembly than the field, enabling faster growth of the FF population compared to field galaxies at fixed stellar mass. This accelerated growth is largely driven by a higher rate of mergers, including systematically more massive mergers, for galaxies inside filaments~(particularly in GAEA2024). As a result, the most massive galaxies are found predominantly within filaments, with their final mass strongly dependent on the time of infall. Galaxies with the highest stellar masses~($\log_{10}(\rm M_{\rm star}/\rm M_{sun}) > 11$) typically have entered filaments more than 9 Gyr ago and remained centrals, thereby avoiding the suppressive effects of being a satellite.  This is also reflected in quenching: the FF population itself exhibits higher quenched fractions than the field in both models, despite never experiencing group processing. We interpret this as evidence of filament-driven quenching, associated with merger activity. The effect is particularly pronounced in GAEA2024.
\par
At the same time, group processing remains important for massive galaxies. The FGF and FFG populations show the highest quenched fractions and the strongest divergence from the field, consistent with the well-known suppressive effects of the group environment. The FF-B population occupies an intermediate regime, suggesting that even transient group interactions can leave a lasting imprint on the properties of massive galaxies. Finally, the TNG100 model predicts a higher fraction of rejuvenated galaxies in filaments compared to the field at fixed stellar mass.
\par
Taken together, our results demonstrate that filaments are not simply transitional pathways to groups and clusters, but environments that directly influence galaxy evolution. Their role is mass-dependent: for low-mass galaxies, groups remain the dominant regulatory mechanism, whereas for massive galaxies, filaments themselves foster accelerated growth, enhanced merger activity, and elevated quenching, even in the absence of group processing. This paper underscores that environmental imprint on galaxies is most evident for long-term filament members, refining the prevailing view of environmental influence and highlighting the need to incorporate full environmental histories, rather than present-day environment alone, when interpreting galaxy evolution.

\begin{acknowledgements}
      .
\end{acknowledgements}

%
  \bibliographystyle{aa} 
  \bibliography{main} 
%
\begin{appendix}

\section{Environmental history variation with adopted filament radii}
\label{appsec:env_hist_var}

\begin{figure}
    \centering
    \includegraphics[width=1\linewidth]{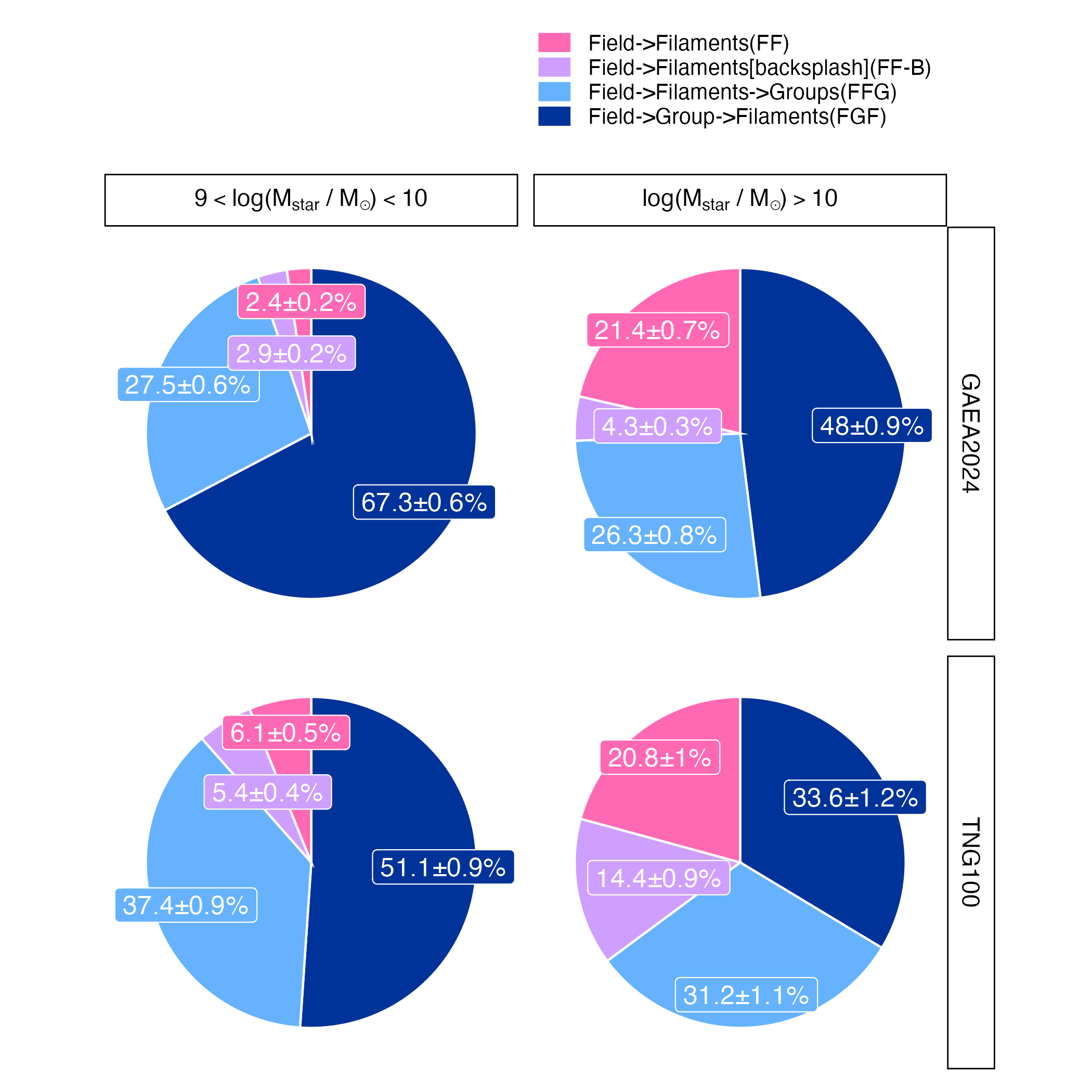}
    \caption{The same as Fig.~\ref{fig:fils_massive_env_history_d1} but when filaments members considered at the distance 0.5 Mpc/h. }
    \label{ffig:fils_massive_env_history_d0.5}
\end{figure}

\begin{figure}
    \centering
    \includegraphics[width=1\linewidth]{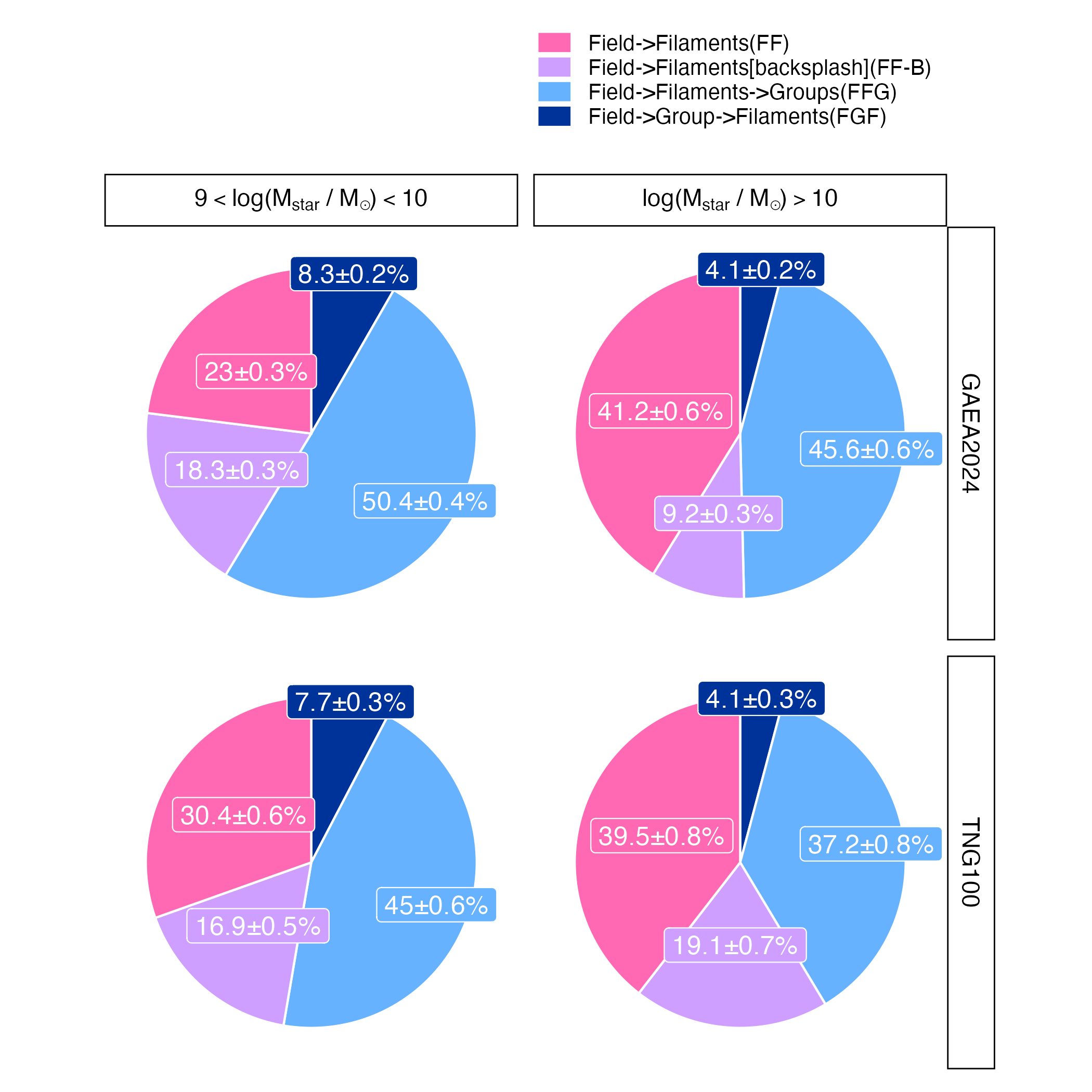}
    \caption{The same as Fig.~\ref{fig:fils_massive_env_history_d1} but when filaments members considered at the distance 2 Mpc/h. }
    \label{ffig:fils_massive_env_history_d2}
\end{figure}

The reported contribution of different environmental histories of galaxies inside filaments depends on the adopted filament radii. In the main analysis, we select all galaxies within 1 Mpc/h of the filament axis, but here we explore alternative choices of 0.5 and 2 Mpc/h. Figures ~\ref{ffig:fils_massive_env_history_d0.5} and \ref{ffig:fils_massive_env_history_d2}  show the fractions of filament members using radii of 0.5 and 2 Mpc/h, respectively. Overall, regardless of the adopted radii, more than half of the population of filament members are galaxies inside groups. When a radius of 0.5 Mpc/h is used, then this fraction is even higher -- reaching around 90-95\% for low-mass and 65-75\% for massive population. This is because galaxy groups within filaments typically lie closer to the filament axis, or because filament-finding algorithms tend to trace the axis through group regions. This is also a reason why the fraction of the FF population increases if we consider bigger filament radii.

\section{Full merger history of FF and field galaxies}
\label{appsec:mergers_full}

\begin{figure*}
    \centering
    \includegraphics[width=1\linewidth]{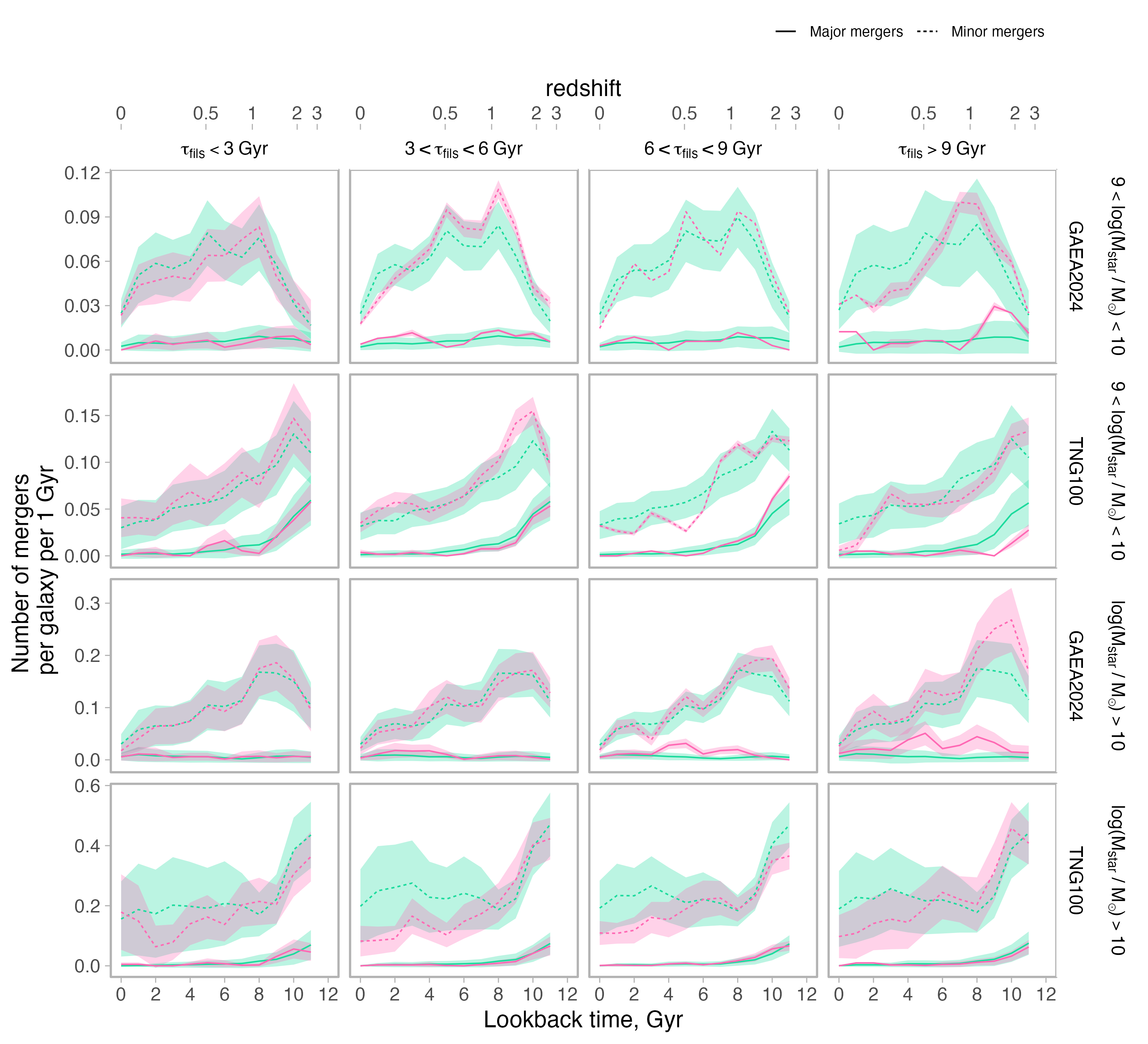}
    \caption{The same as Fig.~\ref{fig:number_of_mergers_major} but for all $\tau_{\rm fils}$ bins.}
    \label{fig:fils_field_merger_rate_full}
\end{figure*}

Here we demonstrate how the merger trends change for the FF population if they infall later than 9 Gyr ago. The latter were discussed in Sec..~\ref{sec:discussion}. Fig.~\ref{fig:fils_field_merger_rate_full} shows all merger rates for the FF and field populations for each of the four $\tau_{\rm fils}$ bins. According to both models, low-mass galaxy merger rates, both major and minor, are similar within the uncertainties between the FF and field populations, regardless of the infall time into filaments. For massive galaxies, TNG100 also predicts the same merger rates regardless of infall time into filaments.
In contrast, the GAEA2024 model predicts that only galaxies that fell early into filaments (> 9 Gyr ago) demonstrate a higher major merger rate compared to the field population when stellar mass is controlled. Moreover, this trend can also be observed in galaxies that infall into filaments between 6 and 9 Gyr ago, although it is weaker. For earlier infall bins, the trend weakens further and becomes undetectable. This reveals the importance of early infall into filaments in uncovering the distinct properties of filaments, according to GAEA2024.

\end{appendix}

\end{document}